\documentclass[lettersize,journal]{IEEEtran}

\usepackage{cite}
\usepackage{amsmath,amssymb,amsfonts}
\usepackage{algorithmic}
\usepackage{graphicx}
\usepackage{textcomp}
\def\BibTeX{{\rm B\kern-.05em{\sc i\kern-.025em b}\kern-.08em
    T\kern-.1667em\lower.7ex\hbox{E}\kern-.125emX}}
\usepackage{amsmath,amsfonts}
\usepackage{algorithmic}
\usepackage{algorithm}
\usepackage{array}
\usepackage{textcomp}
\usepackage{stfloats}
\usepackage{url}
\usepackage{verbatim}
\usepackage{graphicx}
\usepackage{siunitx}
\usepackage{cite}
\usepackage{marvosym}
\usepackage{ragged2e}
\usepackage{pifont}
\usepackage{booktabs}
\usepackage{colortbl}
\usepackage{multirow}
\usepackage{hhline}
\usepackage{vcell}

\usepackage{threeparttable}

\usepackage{array,tabularx,makecell} 

\usepackage{cite}
\usepackage{amsmath,amssymb,amsfonts}
\usepackage{algorithmic}
\usepackage{booktabs}
\usepackage{tabularx}
\usepackage{graphicx}
\usepackage{textcomp}
\usepackage[table,xcdraw,dvipsnames,svgnames,x11names]{xcolor}
\usepackage{multirow}
\usepackage{colortbl}
\usepackage{booktabs}
\usepackage{booktabs}
\usepackage{multirow}
\usepackage{tabularray}

\usepackage[hidelinks,colorlinks=true,linkcolor=Emerald,citecolor=cyan,urlcolor=black]{hyperref}

\newcommand{\Design}{\texttt{\textbf{AMS-HD}}}

\newcommand{\ignore}[1]{ }

\usepackage{tikz,xcolor}

\definecolor{lime}{HTML}{A6CE39}
\DeclareRobustCommand{\orcidicon}{%
	\begin{tikzpicture}
	\draw[lime, fill=lime] (0,0) 
	circle [radius=0.14] 
	node[white] {{\fontfamily{qag}\selectfont \tiny ID}};
	\draw[white, fill=white] (-0.0625,0.095) 
	circle [radius=0.007];
	\end{tikzpicture}
	\hspace{-2mm}
}

\foreach \x in {A, ..., Z}{%
	\expandafter\xdef\csname orcid\x\endcsname{\noexpand\href{https://orcid.org/\csname orcidauthor\x\endcsname}{\noexpand\orcidicon}}
}

\newcommand{\red}{}

\newcommand{\black}{}

\renewcommand{\red}{\textcolor{red}}

\renewcommand{\black}{\textcolor{black}}

\usepackage{tabularx,array}
\newcolumntype{C}{>{\centering\arraybackslash}X} % centered, auto-width

\begin{document}
%\bstctlcite{IEEEexample:BSTcontrol}

\title{
%\huge 
%\textit{AMS-HD}: \underline{A}cute \underline{M}ountain \underline{S}ickness Detection\\ with  \underline{H}yper\underline{d}imensional Computing
AMS-HD: Hyperdimensional Computing for Real-Time and Energy-Efficient Acute Mountain Sickness Detection
}

\author{Abu Masum\orcidA{}, \IEEEmembership{Graduate Member,~IEEE}, Mehran~Moghadam\orcidB{}, \IEEEmembership{Graduate Member,~IEEE},\\ 
 M.~Hassan~Najafi\orcidC{}, \IEEEmembership{Senior Member,~IEEE}, Bige~Unluturk \orcidD{}, \IEEEmembership{Senior Member,~IEEE},\\ Ulkuhan~Guler\orcidE{}, \IEEEmembership{Senior Member,~IEEE}, Beth A. Beidleman \orcidG{}, and Sercan~Aygun\orcidF{}, \IEEEmembership{Senior Member,~IEEE}
\vspace{-2em}
        % <-this % stops a space
\thanks{%\scriptsize{
This work is supported in part by National Science Foundation (NSF) grants 2019511, 2339701, 2609436, in part by National Institute of Health (NIH) under Grant R01HL172293, in part by National Aeronautics and Space Administration (NASA) grant 80NSSC25C0335, in part by Vernon and Ruby Langlinais Non-Endowed Research Fund, in part by the Lockheed Martin Corporation Endowed Professorship Fund, in part by the NASA award, and generous gifts from NVIDIA and Google. A preliminary version of this work appeared as ~\cite{Masum_ISCAS'25}. %was uploaded during the submission of this manuscript.
%in~\cite{Masum_ISCAS'25}.
%}
}
\thanks{%\scriptsize{
Abu Masum and Sercan Aygun are with the School of Computing and Informatics, University of Louisiana at Lafayette, Lafayette, LA 70503, USA. E-mail:\{c00591145, sercan.aygun\}@louisiana.edu. Mehran Moghadam and M. Hassan Najafi are with the Department of Electrical, Computer, and Systems Engineering, Case Western Reserve University, Cleveland, OH 44106, USA. E-mail:\{moghadam, najafi\}@case.edu. Bige Unluturk is with Electrical and Biomedical Engineering, Michigan State University, East Lansing, MI, USA. E-mail:unluturk@msu.edu. Ulkuhan Guler is with Electrical and Computer Engineering Department, Worcester Polytechnic Institute, Worcester, MA, USA. E-mail:uguler@wpi.edu. Beth A. Beidleman is with the US Army Research Institute of Environmental Medicine, Natick, MA, USA. E-mail:beth.a.beidleman.civ@health.mil
\vspace{-0.2em}
}  % <-this % stops a space
%This work is supported in part by the National Science Foundation under Grants No. 2019511, 2339701, and generous gifts from NVIDIA. A preliminary version of this work appeared in~\cite{Masum_ISCAS'25}.}
}
%\markboth{IEEE Transactions on Biomedical Engineering, Vol.~XX, No.~X, XXXX~2026}%
%{Masum \MakeLowercase{\textit{et al.}}: AMS-HD: Hyperdimensional Computing for Real-Time and Energy-Efficient Acute Mountain Sickness Detection}

%\IEEEpubid{0000--0000/00\$00.00~\copyright~2025 IEEE}

\IEEEpubid{
%\vspace{0.5em}
\parbox{\columnwidth}{%\centering
\scriptsize
This work has been submitted to the IEEE for possible publication.
Copyright may be transferred without notice, after which this version
may no longer be accessible.
}
\hspace{\columnsep}\makebox[\columnwidth]{}
}

\maketitle

\begin{abstract}
%\red{[The rule is to have a structured abstract with 5 categories: Objective, Methods, Results, Conclusion, and Significance less than 250 words]}\newline
\color{black}
\textit{Objective:} Acute mountain sickness (AMS) is the most prevalent altitude illness, affecting unacclimatized individuals ascending above 2,500 m and potentially escalating to life-threatening cerebral or pulmonary edema. Conventional machine learning (ML) methods for AMS detection from wearable physiological signals often fail to meet real-time hardware efficiency requirements of continuous monitoring. %This work addresses this challenge.
\textit{Methods:} We present \Design{}, the first \underline{\textbf{h}}yper\underline{\textbf{d}}imensional computing (HDC)-based framework for real-time AMS detection, spanning high-level bipolar (-1/+1) computing for mobile platforms and low-level binary (0/1) computing for FPGA and ASIC targets. The framework integrates mutual information feature selection, hypervector encoding, and positional projection to enhance classification efficiency. Validation spans ARM, FPGA, %an ARM processor, a PYNQ-Z2 FPGA, 
and smartwatch-smartphone platforms using wearable-accessible SpO\textsubscript{2} and heart rate signals.
\textit{Results:} \Design{} matches or outperforms SVM and MLP baselines in both binary and multiclass classification, achieving up to 91\% accuracy and 90\% F1-score in binary classification, and up to 85\% accuracy on external AMS-related datasets. On FPGA, \Design{} reduces LUT and flip-flop usage by 7.3$\times$ and 5.8$\times$, while consuming 3.9$\times$ less power than MLP. On mobile platforms, \Design{} requires only 1\% battery per session, 60 Bytes of memory, and 2.50 ms inference time -- approximately 2$\times$ and more than 3$\times$ lower energy consumption than SVM and MLP. \textit{Conclusion:} \Design{} provides %emerges as 
a scalable, hardware-aware alternative to conventional ML for %continuous and
real-time AMS monitoring, achieving competitive %detection 
performance with substantially lower resource consumption. %across embedded and mobile platforms.
\textit{Significance:} This work presents the first complete HDC framework for %biomedical 
altitude sickness detection, bridging wearable inference and low-level hardware deployment %, and motivating broader adoption of brain-inspired computing 
for resource-constrained health monitoring.
\end{abstract}

\begin{IEEEkeywords}
Acute mountain sickness, hyperdimensional computing, vector symbolic architecture, wearables.
\end{IEEEkeywords}

\vspace{-10pt}

\section{Introduction}
\begin{figure}[t]
%\vspace{-10pt}
\centering
\includegraphics[width=230pt]{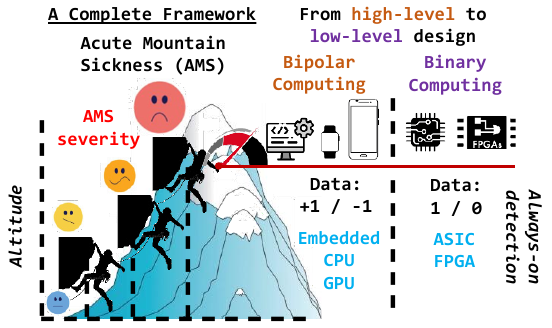}
\vspace{-0.75em}
\caption{Overview of the proposed \Design~framework. The system provides a complete design from high-level bipolar computing ($-1$/$+1$) on mobile and embedded processors to low-level binary computing (logic $0$/$1$) on hardware platforms such as ASICs and FPGAs. By integrating physiological signals with lightweight hyperdimensional operations, the framework enables always-on, resource-efficient detection of acute mountain sickness.}
\label{figure1}
\vspace{-1.5em}
\end{figure}

%Acute mountain sickness (AMS), also referred to as acute altitude sickness (AAS), %(or acute mountain sickness (AMS) as an alternative name) 
\IEEEPARstart{A}{cute} mountain sickness (AMS) is a common medical condition that arises %affecting 
in individuals ascending %who ascend 
\textcolor{black}{to altitudes $\geq$ $2,500$ meters}, where reduced atmospheric oxygen pressure impairs normal physiological function. %the oxygen pressure in the air decreases. 
\textcolor{black}{Early symptoms, including headache, nausea or vomiting, fatigue/weakness, dizziness/lightheadedness, often appear within $6-12$ hours after reaching such altitudes.} If left untreated, %not addressed, 
AMS can progress to severe and potentially life-threatening conditions, such as high-altitude cerebral edema~\cite{hohenhaus1995ventilatory}. Timely detection and intervention are therefore %timely alerting of symptoms are 
critical to reducing mortality. %for mitigating these risks. 
Continuous monitoring of physiological signals using lightweight wearable devices offers a promising approach~\cite{costanzo_rbme_2022}, as rapid analysis and early diagnosis of symptoms can enable timely intervention and substantially lower the risk of fatal outcomes.
%A promising solution for continuous monitoring of vital signals is the use of lightweight wearable devices. Early diagnosis through rapid symptom analysis can significantly reduce fatalities.
Traditional machine learning (ML) techniques have been investigated for \textcolor{black}{AMS detection~\cite{wei2021using,li2025acute, wang2024recent}}, but their reliance on substantial computational resources limits their suitability for real-time diagnostics in wearables.
%~\cite{chen2024machine}. 
%However, these methods often rely on substantial computational resources and are not well-suited for real-time diagnostics in portable devices. 
To address this challenge, this work proposes, for the first time, an alternative solution for fast, accurate, and resource-efficient detection of AMS by leveraging an emerging model of computing, \textit{hyperdimensional computing (HDC)}. We present a complete framework that spans both high-level bipolar computing ($-1$/$+1$), suitable for mobile platforms, and low-level binary computing (logic $0$/$1$) optimized for FPGAs and ASIC implementations (Fig.~\ref{figure1}). %application-specific digital circuit designs (Fig.~\ref{figure1}).

HDC is an emerging brain-inspired paradigm that seeks to replicate aspects of human memory, perception, and cognition through mathematical operations in high-dimensional spaces~\cite{kanerva2009hyperdimensional}. In HDC, data are represented as \textit{hypervectors} ($\mathcal{HV}$s)--one-dimensional vectors consisting of hundreds to thousands of random binary (`0's and `1's) or bipolar (`$-1$'s and `$+1$'s) elements. These $\mathcal{HV}$s are designed to be (nearly) orthogonal, ensuring independence between representations and enabling robust, noise-tolerant computation. Unlike conventional ML models that rely on complex arithmetic, HDC performs learning and inference through highly parallelizable, hardware-efficient vector operations such as permutation, binding, and bundling~\cite{rasanen2014modeling}. This enables HDC to achieve fast, resilient performance across diverse tasks, while requiring minimal memory and energy resources~\cite{aygun2023linear}. Recent advances in $\mathcal{HV}$s generation, including low-discrepancy (LD)~\cite{aygun2024sobol, shoushtari2024all, aygun2024uhd,ID_VSA_Moghadam_TVLSI2026}
and Hadamard-based~\cite{masum2025fly} encodings, have further enhanced orthogonality and accuracy, drawing inspiration from quasi-randomness techniques used in stochastic computing (SC)~\cite{8327916,Moghadam_Robust_GLSVLSI25}. Together, these properties make HDC a compelling alternative to traditional ML approaches for real-time, resource-constrained biomedical applications~\cite{Phoneme_Recog_HDC_TBME_2024,Seizure_HDC_TBME_2020}.

\begin{figure}[t]
%\vspace{-10pt}
\centering
\includegraphics[width=190pt]{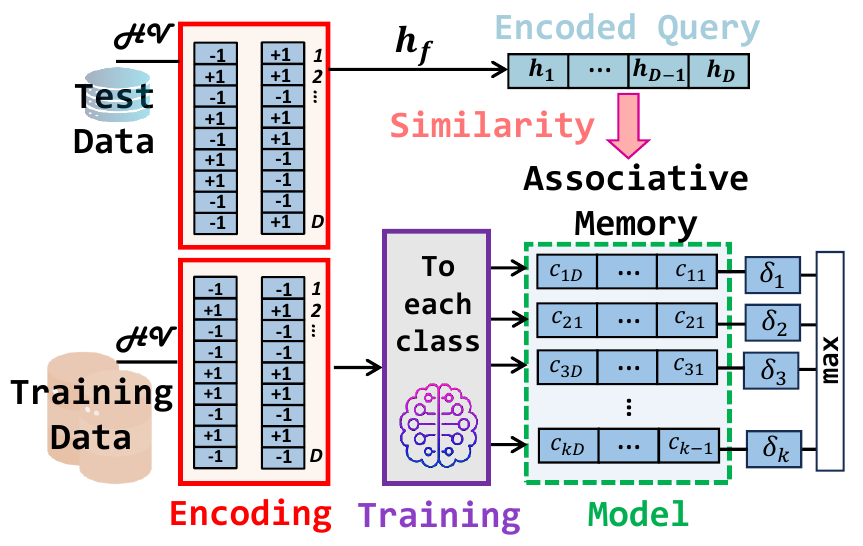}
\vspace{-1em}
\caption{Overview of an HDC model: encoding, training, and classification via similarity search.}
\label{fig:vsa_fig1}
\vspace{-1em}
\end{figure}

Fig.~\ref{fig:vsa_fig1} illustrates the general workflow of an HDC model. Input features from training and test data are encoded into high-dimensional $\mathcal{HV}$s through lightweight logic operations (e.g., \texttt{XOR}, \texttt{Add}, \texttt{Shift}, and \texttt{Permute}). During training, each encoded sample incrementally contributes to class representations, forming a distinct $\mathcal{HV}$ class for every category. This process yields the final deployable model. %, i.e., associative memory}. 
In inference, test samples are encoded in the same manner to generate query $\mathcal{HV}$s. %which are then 
The query $\mathcal{HV}$ is then compared against all stored class \textit{$\mathcal{HV}$s} using similarity metrics, and the class with the highest similarity score is identified as the prediction.

Building on this foundation, we introduce~\Design, a vector symbolic architecture tailored for AMS detection. The framework integrates optimized feature encoding and circadian-aware physiological signals collected from wearable devices, enabling real-time monitoring of vital signs. Extensive evaluations on FPGA prototypes and mobile-wearable device %Android-based
implementations confirm the practicality of \Design~across diverse hardware platforms, highlighting its suitability for resource-constrained environments. By uniting hardware-efficient design with predictive accuracy, \Design~demonstrates the transformative potential of HDC in biomedical applications, paving the way for scalable, always-on health monitoring and proactive management of high-altitude illness.  The key contributions of this study
are as follows:\\
\ding{182} We introduce \Design, the first HDC-based framework specifically designed for real-time detection of AMS on wearable and mobile platforms.\\
\ding{183} The framework encompasses high-level bipolar computing ($-1$/$+1$) for mobile applications and low-level binary computing (0/1) for efficient deployment on hardware platforms, supporting circuit-level design. Notably, bipolar computing brings HDC closer to real-world applications that run directly on mobile devices.\\
\ding{184} We propose optimized feature encoding strategies in the preprocessing step to improve %robustness,
detection accuracy and energy efficiency.\\
\ding{185} We validate \Design~across multiple platforms: an FPGA prototype, an \textcolor{black}{Advanced RISC Machine (ARM)-based} embedded processor, and a %mobile Android
smartwatch–mobile system, demonstrating low latency, low energy consumption, and suitability for continuous monitoring.

%\color{black}
%The remainder of this manuscript is organized as follows. Section~\ref{background} discusses the background on HDC and traditional learning methods for AMS prediction. Section~\ref{Sec:ProposedDesign} discusses the proposed~\Design~framework. Section~\ref{evaluation} provides the experimental results, including dataset and %, evaluation metrics, and
%platform configurations, and evaluation results. Section~\ref{Sec:SOTA} provides an extensive comparison with the state-of-the-art (SOTA). Finally, Section~\ref{sec:conclusion} concludes the paper.

\section{Background} \label{background}

\subsection{An Emerging Computing Paradigm: HDC (a.k.a. VSA)} 
HDC is an emerging computational model that enables efficient and lightweight data processing. It has demonstrated effectiveness across diverse applications, including classification, pattern recognition, cognitive modeling, and learning with limited data~\cite{kleykoSurvey,Seizure_HDC_TBME_2020}. Its key strengths include rapid training and suitability for low-power hardware, making it particularly attractive for real-time and resource-constrained systems.
One of HDC's key strengths is its ability to represent symbolic information through nearly orthogonal structures, which forms the basis of  %this allows it to create 
vector symbolic architectures (VSAs). 
VSAs can represent not only scalar values but also symbolic information (e.g., positions, characters, signal timestamps, etc.) in a high-dimensional binary vector format~\cite{RAHIMI2020195}. %enabling holistic data 
VSAs %Symbolic architectures 
exhibit strong robustness to noise and errors, owing to their distributed data representation. Unlike conventional binary formats, their performance does not rely on the significance of individual bits (e.g., the least significant bit, the most significant bit, or the sign bit).

\begin{figure*}[!t]
\centering
%\vspace{-10pt}
\includegraphics[width=\linewidth]{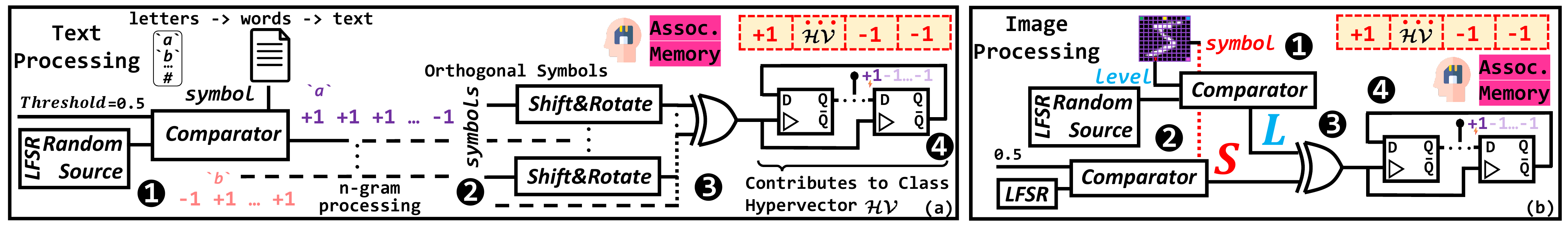}
\vspace{-2em}
\caption{Neuro-symbolic learning architectures and their corresponding encoding: (a) language text processing using n-gram encoding, and (b) image processing using record-based encoding (+1s represent logic-1s in memory, -1s represent logic-0s in memory.)  
}
\vspace{-1em} 
\label{VSA_general} 
\end{figure*}

In HDC, %HDC-based VSA \red{[MM-Does it make sense? What is HDC-based VSA? HDC and VSA are different terminologies for the same concept.]}, 
high-dimensional binary (or bipolar) $\mathcal{HV}$s are generated such that they exhibit significant dissimilarity from one another, allowing unique symbols to be distinguished with high fidelity. In neuro-symbolic learning, orthogonality is critical for accurate learning and recognition, while the distributed representation enhances resilience to noise and errors, making HDC particularly robust for diverse learning tasks~\cite{imani2019framework, RahimiHamming}. Fig.~\ref{VSA_general} illustrates two common neuro-symbolic learning architectures: n-gram- and record-based-learning architectures for text and image processing, respectively.

$\mathcal{HV}$s are generated by comparing random source values with positional indicators %(i.e., 0.5 unbiased probability between $P=0$ and $P=1$) 
or with scalar values over $D$ cycles. The encoding process then applies a sequence of lightweight logical operations, including \textit{permutation} (Step-2 in Fig.~\ref{VSA_general}(a), only for record-based encoding), \textit{binding} (Step-3 in both Fig.~\ref{VSA_general}(a) and (b)), and \textit{bundling} (Step 4 in both cases)~\cite{Kazemi2022}. Permutation reorders vector elements to preserve orthogonality\cite{rahimi2016robust}, binding combines multiple $\mathcal{HV}$s through element-wise multiplication of bipolar values ($\pm 1$)—or \texttt{XOR} in the binary computing case—and bundling merges multiple $\mathcal{HV}$s into a single composite representation while maintaining data integrity. Traditionally, $\mathcal{HV}$s are generated using pseudo-random methods, such as using Linear-Feedback Shift Register (LFSR) hardware, which may lead to limited orthogonality and degraded performance~\cite{shoushtari2024all}. To overcome this limitation, quasi-random sequences such as Sobol have been introduced~\cite{aygun2023linear,Mehran-No-Mult2023}, improving orthogonality and enhancing the reliability of neuro-symbolic encodings.

\subsection{AMS Prediction: Conventional ML Approaches} 
ML algorithms have been widely investigated for predicting the severity of AMS, enabling timely interventions such as descent from high-altitude environments.
Prior studies have applied both regression- and classification-based methods using physiological and environmental data as predictors~\cite{wu2020assessment, beidleman2017predicting}. For example, Yang et al.\cite{YangETAL} employed ML techniques to assess susceptibility to severe AMS (sAMS) using genetic information; however, the real-time application of genetic data is computationally challenging and impractical for portable systems. In contrast, physiological signals such as blood oxygen saturation (\textit{SpO\textsubscript{2}}) and heart rate (\textit{HR}) provide more accessible and informative features for AMS prediction\cite{dunnwald2021use}. Additional contextual variables, including altitude, relative humidity, climbing speed, and heart rate variability have also been shown to improve prediction accuracy~\cite{wu2020assessment,dunnwald2021use}. A variety of ML models have been explored for this task, such as bagged trees, logistic regression (LR), linear support vector machines (SVM), and weighted k-nearest neighbors (kNN)\cite{wei2021using}. Beyond these classical approaches, recent studies have highlighted links between sleep quality and AMS, suggesting that circadian rhythm and intermittent hypoxia (alternating between normal oxygen levels and low oxygen levels) training may also influence susceptibility. For instance, Wang et al.\cite{wang2023slow,wang2024event} employed pseudo-labeling techniques and trained long short-term memory (LSTM) networks~\cite{greff2016lstm} to classify hypoxia tolerance, reporting promising results. While such ML solutions demonstrate strong predictive performance, they generally require powerful computing platforms, multiple layers of optimization, and high resource consumption, making them unsuitable for real-time deployment on wearable or embedded devices. In contrast, HDC, with its neuro-symbolic learning capabilities, offers lightweight and fast processing, supporting single or few-shot learning. HDC presents low power and low memory usage~\cite{10.1145/3649476.3658795}, making it a promising candidate for resource-constrained environments such as wearable health monitoring systems~\cite{Sleep_Apnea_HDC_TBME_2024}.

\section{Proposed Framework}
\label{Sec:ProposedDesign}
%This section presents a detailed explanation of the methodology of our classifier. The overall framework is given in Fig.~\ref{fig:proposed_method}.

This section presents the proposed framework. We describe how our HDC classifier is applied to classify the AMS dataset using physiological features. %The inputs include oxygen saturation (SpO\textsubscript{2}), heart 
%As illustrated in Fig.~\ref{fig:proposed_method},~\Design~%employs HDC to 
The proposed~\Design~framework performs both binary and multiclass AMS classification. Our study emphasizes feature engineering, systematic data analysis, and the evaluation of HDC for lightweight system design. Each stage of the system, from raw data to the HDC classifier, is described in detail. We use biomedical tabular data, where patient samples are recorded at varying altitudes with associated AMS scores. %Below, we first describe the dataset. 

\vspace{-0.75em}
\subsection{Dataset Overview}
In this study, we utilize a publicly available dataset\footnote{Dryad repository: \href{https://datadryad.org/dataset/doi:10.6086/D1XM45}{doi:10.6086/D1XM45}} by Pham et al., comprising physiological parameters~\cite{pham2022inflammatory,roach20182018}. The dataset captures the body’s response to high-altitude stress, with a %particular 
focus on immune system changes.
It includes key physiological features such as oxygen saturation (SpO\textsubscript{2}), heart rate (HR), carbon monoxide (CO) measured %both 
in percentage (\%) and parts per million (ppm), systolic and diastolic blood pressure (Psys and Pdia), hematocrit (Hct), and \textcolor{black}{Lake Louise-referenced AMS score~\cite{roach20182018}. 
The dataset comprises both wearable-accessible and blood-sample-derived variables; however, the blood-derived variables are not intended for real-time deployment. In this work, SpO\textsubscript{2}, HR, and event/time information are considered wearable- or mobile-accessible inputs. In contrast, Hct and CO-related measurements are blood-sample-derived variables; they are included for dataset characterization and feature-importance comparison, but they are not required %inputs 
for deployment-oriented \Design{} learning or a real-time inference pipeline and are not used in the training or inference pipeline in this work. Among the wearable-accessible features, SpO\textsubscript{2} represents the percentage of oxygen saturation in the blood, while HR measures the number of heartbeats per minute. Among the blood-derived features, CO (\%) and CO (ppm) indicate the presence of carbon monoxide in the blood, which can exacerbate oxygen deprivation %in the body 
at higher altitudes. Hematocrit (Hct) refers to the proportion of red blood cells in the blood.}

\textcolor{black}{The dataset by Pham et al.~\cite{pham2022inflammatory} includes a protocol-level \textit{Event Name} field and a corresponding ordered \textit{Time} field, which together identify the measurement stage of each physiological record. Specifically, measurements are organized across exposure stages such as \textit{Baseline (sea level)}, \textit{Night~1--3 (altitude)}, and \textit{Overnight~1--3}. In this dataset, the \textit{night} labels denote measurements collected during nighttime altitude exposure, whereas the \textit{overnight} labels refer to measurements collected after a full night of exposure. The %corresponding 
\textit{Time} values provide an ordered stage index (e.g., 0, 0.5, 1, 1.5, 2, 2.5, and 3), reflecting the progression of the measurement protocol. Thus, these event-time numerical variables are used as contextual stage information associated with each physiological measurement, rather than as dense %continuous 
wearable time-series streams or categorical labels.} 

\begin{figure*}[!t]
    \centering
    %\vspace{-10pt}
    \includegraphics[width=\linewidth]{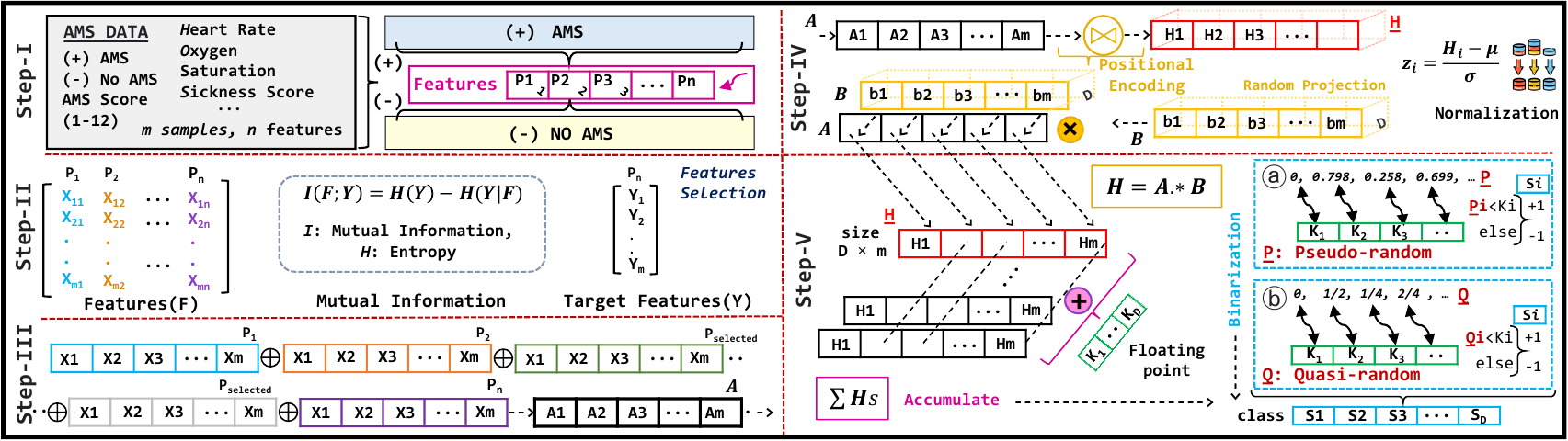}
    %\vspace{-10pt}
    \caption{{Proposed}~\Design~training pipeline: MI feature selection, positional encoding, hyperdimensional projection, and \textcircled{a} pseudo- vs. \textcircled{b} quasi-random binarization.
    %\Design:~Proposed framework for classification task: Positional Encoding, Random $\mathcal{HV}$ Projection, and Class Vector (with \textcircled{a} Pseudo- vs. \textcircled{b} Quasi-randomness.
    %vs \red{\textcircled{c} Hadamard})
    }
    \label{fig:proposed_method}
    \vspace{-0.5em}
\end{figure*}

\textcolor{black}{The dataset also includes the Lake Louise-referenced AMS scores~\cite{roach20182018}, ranging from 0 to 12, which quantify the severity of AMS symptoms. According to~\cite{roach20182018}, an AMS score $\geq$ 3 \textit{together with headache} indicates AMS. %that a patient is sick .
Based on the available AMS scores, we evaluate two classification settings. For binary classification (\texttt{Step-I} in Fig.~\ref{fig:proposed_method}), samples with AMS scores 0--2 are labeled as \texttt{No AMS}, while samples with scores 3--12 are labeled as \texttt{AMS}. Because the dataset by Pham et al.~\cite{pham2022inflammatory} used in this work does not include headache information, these binary labels should be interpreted as score-based AMS labels rather than full clinical diagnoses requiring headache confirmation. For multiclass classification, AMS scores are grouped into four severity levels: \textit{No AMS} (0--2), \textit{Mild AMS} (3--5), \textit{Moderate AMS} (6--9), and \textit{Severe AMS} (10--12), enabling finer-grained modeling of symptom severity based on the available score values. Because the original dataset has limited and imbalanced samples across severity levels, especially for moderate and severe AMS, we augment only the training data to reduce class bias while preserving the same score-based label definitions.}

%\textcolor{red}{The AMS score, ranging from 0 to 12, quantifies the severity of AMS symptoms (Lake Louise 2018~\cite{roach20182018}). The AMS score $\geq$ 3 \textit{plus headache} indicates that a patient is sick.} 

\subsection{Data Preparation and Analysis}
\begin{figure}[t]
%\vspace{-10pt}
    \centering
    \includegraphics[width=\linewidth]{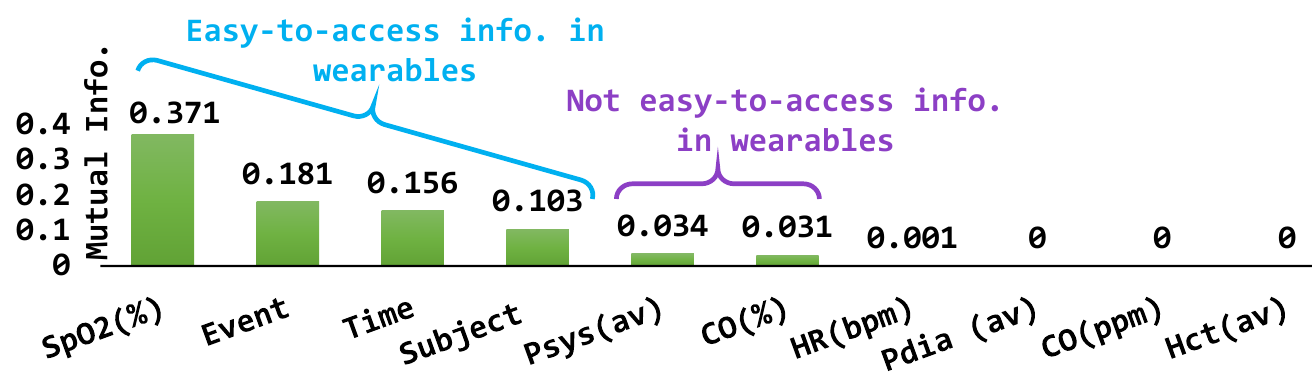}
    \vspace{-1.5em}
    \caption{Feature importance scores calculated via MI, highlighting SpO$_2$ (\%) as the most influential feature.}
    \label{fig:features}
    \vspace{-10pt}
\end{figure}

To ensure the dataset's integrity and suitability for analysis, several preprocessing steps were implemented, including handling missing values, feature selection, and normalization. Missing values, which could introduce bias and affect model performance, were imputed using the mean of the respective feature across all subjects, preserving the dataset's statistical properties. The dataset, structured by subject IDs, includes corresponding event-time numericals and physiological parameters. To assess the risk of AMS, we analyzed variations in physiological responses as a function of environmental factors such as event type and time. By integrating these contextual variables with physiological measurements recorded at different altitudes and time intervals, we observed that a rapid decline in SpO\textsubscript{2} levels or an increase in heart rate, key physiological indicators of AMS, was strongly correlated with altitude gain. \textcolor{black}{Fig.~\ref{fig:proposed_method} should be interpreted as the dataset-driven, sample-level \Design{} training pipeline used in the current evaluation. \texttt{Step-I} summarizes the available AMS-related input attributes and score-based labels. In this pipeline, the event-time variables serve as contextual attributes for each physiological measurement, reflecting conditions such as baseline, sea-level, high-altitude, night, and overnight stages. These variables are not encoded via a separate temporal-permutation mechanism in the current evaluation, because the available dataset provides discrete event-time annotations rather than dense, continuous wearable streams. However, in the always-on monitoring design, explicit permutation-based temporal encoding can be incorporated as a future extension to capture temporal dynamics and physiological progression from continuous wearable data.} We apply feature selection as part of our feature engineering pipeline. As shown in \texttt{Step-II} of Fig.~\ref{fig:proposed_method}, we utilize feature selection using  
Mutual Information (MI) to evaluate the relevance of each feature ($F$) with respect to the target attribute ($Y$). MI quantifies how much information a feature contributes to predicting the target variable. Our dataset consists of $m$ samples and $n$ features, where the target feature (AMS Score) is denoted as $Y$, and the feature matrix $F$ contains the remaining features. The conventional definition of MI is given as $I(F; Y) = H(Y) - H(Y|F)$, where $H(Y)$ denotes the entropy of the AMS score, and $H(Y \mid F)$ represents the conditional entropy of $Y$ given $F$. %measures the uncertainty of $Y$ after knowing $F$.
Since the dataset contains continuous physiological parameters, entropy cannot be directly computed. Instead, MI is estimated using a k-nearest neighbors (kNN)-based method, which captures %determines MI by analyzing 
local density variations in the joint feature-target space $(F, Y)$, and is expressed as: 
\vspace{-0.5em}
\begin{equation}
I(F; Y) \approx \psi(N) - \psi(N_{x_i}) + \psi(k) - \psi(m_i)
\end{equation}

\noindent where $\psi(N)$ is the total number of samples, $\psi(N_{x_i})$ the neighbors in the feature distribution, $\psi(k)$ the nearest neighbors, and $\psi(m_i)$  the neighbors within the k$^{th}$ nearest neighbor's distance~\cite{ross2014mutual}. A higher MI value indicates that feature $F$ provides stronger predictive power for %significantly predicts 
the target variable $Y$, whereas a lower value suggests limited relevance. %the opposite. 
\texttt{Step-III} in Fig.~\ref{fig:proposed_method}  demonstrates the combination of the selected features from different sensors through point-wise accumulations. \textcolor{black}{To clarify the deployment feasibility of these features, Fig.~\ref{fig:features} separates features that are easy to access in wearable or mobile settings from those that require invasive or laboratory-based measurements. This distinction is important because the complete dataset includes both real-time wearable-accessible signals and blood-sample-derived variables.} Fig.~\ref{fig:features} illustrates the MI feature importance values obtained from offline dataset analysis, indicating the degree of association between each feature and the target variable in the classification task. \textcolor{black}{The results show that, %from the dataset demonstrate that, 
among all features, SpO\textsubscript{2} (\%) has the strongest influence, with an MI score of $0.371$. Features such as CO (\%) ($0.034$) and HR (bpm) ($0.031$) contribute less, while others, including Pdia (av), CO (ppm), and Hct (av), receive an MI score of $0$, suggesting minimal or no direct contribution to classification in this dataset. %Nevertheless, despite their low or negligible MI values, all features are retained in the model due to their potential medical significance and underlying physiological relationships, which may not be fully captured by MI alone. 
These results further support the wearable-oriented design: the strongest predictor, SpO\textsubscript{2}, is directly measurable using current wearable or pulse-oximetry sensors, whereas blood-sample-derived features contribute relatively little in this dataset. Although HR has a lower MI score, it is retained in the deployment-oriented feature set together with SpO\textsubscript{2}, because it is non-invasive, directly available from current smartwatch sensors, and physiologically relevant for AMS monitoring. Event-Time information is also retained as protocol-level contextual information, while blood-sample-derived variables are excluded from the deployment-oriented \Design{} learning and real-time inference pipeline.}

\subsection{Positional Encoding and Data Projection}
The selected features are positionally (symbolically) encoded. This step is followed by a point-wise multiplication that projects the positions. The $D$ dimensionality is adjusted using a random matrix $B$ in \texttt{Step-IV}, which represents the \textit{binding} operation of the HDC system. Each positionally encoded vector provides an enhanced representation of the input feature(s) based on point-wise position. The positionally encoded values are normalized according to $z$-score normalization with the $B$ random projection matrix: $z_i = \frac{E_i - \mu}{\sigma}$, where \( \mu \) and \( \sigma \) denote the \textit{mean} and \textit{standard deviation} of the features, respectively. The resultant, $H$, contains $m$ selected features of $n$ samples from a patient. Multiple patients of the same class ($H$s) (e.g., no-AMS patients) contribute the same \textit{class} $\mathcal{HV}$ via accumulation.

\subsection{Generating Class $\mathcal{HV}$s}
\texttt{Step-V} in Fig.~\ref{fig:proposed_method} shows the process of generating non-binary scalars or floating-point $\mathcal{HV}$s, %using random $\mathcal{HV}$s, 
using \( H\in\mathbb{R}^{D \times m} \). Different $H$s coming from the patients are accumulated through %, following 
a process analogous to \textit{bundling} of HDC systems. The resulting $\mathcal{HV}$s in floating points can reach dimensionalities as large as \( D \) =  10,000, enabling the embedding of feature vectors into a high-dimensional space.
%that captures %providing the advantage of capturing 
%complex patterns and enhances dataset representation. % within the dataset for holistic representations. 
Following \textit{bundling}, the generated scalar vectors ($K$s) are converted into the binary/bipolar domain to produce %This conversion leads to the formation of 
the class $\mathcal{HV}$s ($S1$,...,$S_D$), which are critical %step to further 
for subsequent high-dimensional processing. This conversion can be performed %is achieved 
using %either 
\textcircled{a} pseudo-random or \textcircled{b} quasi-random approaches.

%, or a novel approach that we propose in this study based on Hadamard sources~\cite{shlichta1979higher}.

\subsubsection{\textcircled{a} Pseudo-random $\mathcal{HV}$ Generation (HDC-P)}
Pseudo-random $\mathcal{HV}$s (\texttt{Step-V} \textcircled{a} in Fig.~\ref{fig:proposed_method}) are generated using pseudo-random number sources derived either from hardware components, such as LFSRs, or from software platforms, such as Python's \texttt{rand()} or MATLAB's \textit{random} functions. Alternatively, these %pseudo-random numbers 
can be derived from %a set of 
predefined random sequences %which can be 
specifically designed to maintain randomness and preserve statistical properties. The resulting $\mathcal{HV}$s are pseudo-orthogonal, meaning that $\mathcal{HV}$s from different classes remain largely distinct %are separated and distinct 
and well-separated in the high-dimensional space. However, achieving optimal $\mathcal{HV}$s for accurate classification requires multiple training iterations to ensure sufficient orthogonality and improve classification performance.

\begin{figure*}[!t]
    \centering
    %\vspace{-10pt}
    \includegraphics[width=0.9\linewidth]{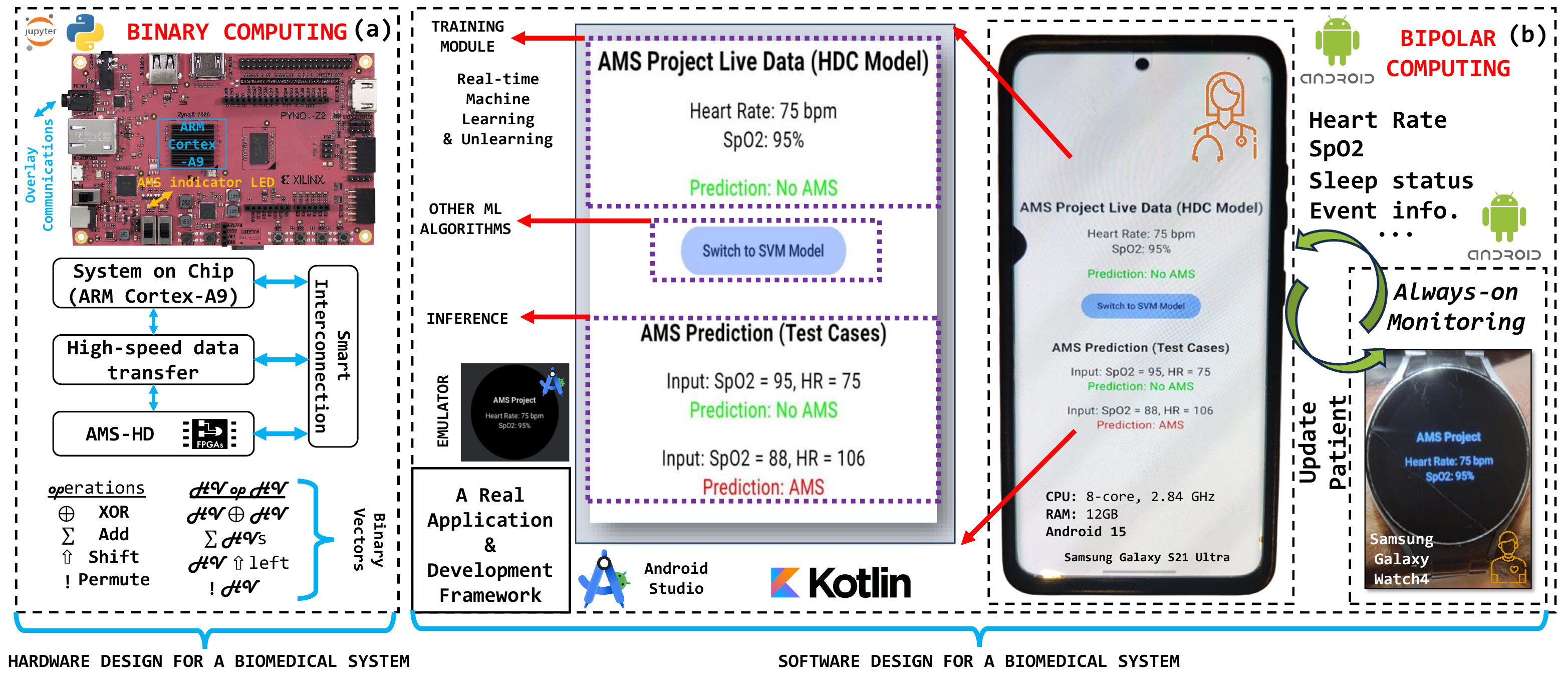}
    \vspace{-0.5em}
    \caption{A real application on the Mobile Phone \& Smart Watch, including the development framework. AMS: Acute Mountain Sickness case study.}
    \label{fig:app}
    %\vspace{-5pt}
    %\red{[Hassan: Since we are submitting to a Circuit and System transaction, they like to see the primarily focus on hardware and circuit. I like this figure (Fig. 9) but still suggest we add a separate figure showing Mehran's FPGA in  large size (similar to the one we had on the paper but removed). We don't want they feel we have developed a sofware-mobile platform. ]}
    \vspace{-10pt}
\end{figure*}

\subsubsection{\textcircled{b} Quasi-random $\mathcal{HV}$ Generation (HDC-Q)}
Quasi-random $\mathcal{HV}$s (\texttt{Step-V} \textcircled{b} in Fig.~\ref{fig:proposed_method}) are generated using quasi-random sequences, such as Sobol~\cite{Najafi_TVLSI_2019}. Unlike pseudo-random sequences, these sequences exhibit the low discrepancy (LD) property, characterized by uniformly distributed points and recurring patterns. This structured randomness produces $\mathcal{HV}$s that are more orthogonal and better distributed compared to those generated with pseudo-random methods. As a result, quasi-random $\mathcal{HV}$s yield higher-quality vector representations, improved class separation, and enhanced overall classification performance. 

In this work, we also explore an alternative approach by employing \textit{Hadamard} matrices for the source of randomness when encoding the positions. %, which are square matrices containing only $\pm1$ values and mutually orthogonal rows
A Hadamard matrix $h_n$ of order $n$ is a square matrix whose rows are mutually orthogonal~\cite{shlichta1979higher,craigen2006hadamard,horadam2012hadamard,butson1962generalized, hedayat1978hadamard}, satisfying $h_nh_n^T = nI_n$. Each row of the Hadamard matrix can be used as a Hadamard sequence. The structured orthogonality of Hadamard rows enables them to serve as deterministic $\mathcal{HV}$s, making them especially suitable for symbol representations (e.g., positions) that clearly separate dimensions while requiring only a single bit per element for storage. Their simple binary structure also enables efficient computation: complex multiplications can be replaced by fast bitwise operations such as \texttt{XOR}, significantly reducing power consumption and latency~\cite{pratt1969hadamard}. These properties make Hadamard $\mathcal{HV}$s particularly attractive for high-speed, energy-efficient implementations of HDC. 
%in wearable and embedded biomedical systems.

\subsection{HDC Classifier %and Usage of 
%Hadamard Matrix
}
The HDC classifier constitutes the final model, obtained by aggregating the contributions from each class $\mathcal{HV}$ ($S$). The binary classifier distinguishes between ``\texttt{AMS}" and ``\texttt{NO AMS}". During inference, test input $\mathcal{HV}$ is compared against trained class $\mathcal{HV}$s ($S_i$ = \texttt{AMS} $\mathcal{HV}$ or $S_i$ = \texttt{NO AMS} $\mathcal{HV}$). Classification is performed using similarity metrics, 
%he classification relies on binary similarity metrics, 
such as %using the binary similarity: E.g., 
cosine similarity: $(\mathcal{HV}, S_i) = \frac{\mathcal{HV} \cdot S_i}{\| \mathcal{HV} \| \| S_i \|}$. This yields 0 score for orthogonal $\mathcal{HV}$s. Similarly Hamming distance: $Hamming(\mathcal{HV},S_i)=\frac{1}{D}\sum_{i=1}^{D}(\mathcal{HV}\!\oplus\! S_i)$ can also be utilized, which gives 0.5 score for orthogonal $\mathcal{HV}$s; $\oplus$ denotes element-wise \texttt{XOR}. After comparing the test vector with the class vectors, the class with the highest similarity score is selected as the prediction.

%\subsubsection{\textcircled{c} 
%Hadamard $\mathcal{HV}$ Generation (HDC-H)}

%\red{[Hassan: The following paragraph is not yet connected and the reader get lost why we are discussing this after HDC similarity check. In current shape, it's better to move it back to the HV generation subsection after discussing Quasi-random generation.]}

\subsection{Processing Platforms: Binary \& Bipolar Computing}
%\subsection{Processing Platforms: Binary \& Bipolar Computing}

The processing platforms of the proposed~\Design~framework are shown in Fig.~\ref{fig:app}: 
(a) a hardware path implementing the HDC classifier with binary computing on an ARM Cortex-A9 SoC, %with an FPGA overlay, 
where feature and position $\mathcal{HV}$s are generated via FPGA design and encoding is completed by lightweight circuits, with an LED indicating AMS status; and 
(b) a bipolar computing path on a mobile-watch device pair.

Similar to the binary design, the mobile system is built on our HDC model and is optimized for adaptive, efficient AMS prediction in mountaineers using always-on smartwatches for continuous sensing. A Samsung Galaxy Watch4 continuously records HR (bpm) and SpO$_2$ and streams the data to a paired smartphone for AMS risk prediction. The platform also supports switching between alternative ML models for comparison. Developed in Android Studio/Kotlin, the application is lightweight and integrates seamlessly with mobile devices. 
\textcolor{black}{For the mobile-wearable deployment path, our real-time sensing pipeline uses wearable-accessible signals, primarily SpO$_2$ and HR, streamed from the smartwatch to the smartphone. Blood-sample-derived variables are not required for real-time mobile inference.}

%The processing platforms of the proposed~\Design~framework are shown in Fig.~\ref{fig:app}. Fig.~\ref{fig:app}~(a) depicts the hardware path used to implement the HDC classifier with binary computing on an ARM Cortex-A9 SoC with an FPGA overlay. The design generates feature and position HVs on-chip, and completes the encoding operations over lightweight circuits, with an indicator LED signaling AMS status. Fig.~\ref{fig:app}~(b) presents the bipolar computing path on an Android phone–watch pair. 

\ignore{
Like binary design, the mobile system is built on our proposed HDC model. %which is 
This high-level system is optimized for adaptive and efficient AMS prediction in %considering the
mountaineers using always-on smart watches for continuous sensing. %to collect the sensor data. 
The wristband continuously records real-time physiological signals, including HR (bpm) and SpO$_2$, via a Samsung Galaxy Watch4, and communicates the data to a paired smartphone for AMS risk prediction. %in coordination with a mobile phone. 
The platform also offers flexibility to switch between alternative ML models for comparison. Developed using Android Studio and Kotlin, our application is lightweight and designed for seamless integration into mobile devices.
}

The architecture supports real-time inference, learning, and unlearning directly on the mobile platform, without reliance on cloud connectivity. The HDC model performs prediction using lightweight vector operations, delivering outputs such as (``Prediction: \texttt{AMS}'') or absence (``Prediction: \texttt{No AMS}''). When a potential risk is detected, the system immediately triggers an alert, enabling timely intervention. To enhance versatility, the platform allows switching between %provides an option to switch between 
models, for example, %enabling comparative analysis by 
running an SVM-based classifier in place of the HDC model. This facilitates comparative analysis %modularity supports the evaluation of alternative algorithms under the same 
under a unified application interface. In addition, the application includes a test-case validation module that enables input simulation for %where input values can be simulated for 
inference testing and debugging. The wearable interface displays key physiological indicators live, HR and SpO\textsubscript{2}, which update in sync with the paired mobile device to provide consistent feedback to the user. Computational performance analysis confirms that the application operates efficiently even on mainstream mobile hardware (e.g., smartphones with 8-core CPUs and 12 GB RAM, running Android 15). By analyzing RAM and CPU usage, we validate that the system maintains low overhead and fast response times. 

\black{Beyond the current real-time inference capability, this always-on wearable platform also provides a natural path for explicit temporal modeling. In the available dataset, event-time information is provided as contextual annotations rather than as dense continuous wearable streams. Therefore, the current \Design{} evaluation uses event-time attributes as contextual features associated with each physiological measurement. In a real-time deployment with continuous smartwatch streams, however, consecutive SpO\textsubscript{2} and HR sample $\mathcal{HV}$s can be temporally permuted and bundled to encode physiological trajectories. This would allow the system to capture unit-time physiological changes and sustained AMS progression, potentially reducing transient false positives. Since HDC permutation can be implemented as a lightweight shift-and-rotate operation, this extension can be incorporated with minimal computational overhead in future continuous monitoring designs.}

%This case study validates the effectiveness of the proposed~\Design~in delivering real-time, energy-efficient AMS detection, underscoring its viability for mobile and wearable healthcare applications, especially in remote or high-risk environments.

%Developed using Android Studio and Kotlin, the application is lightweight and optimized for deployment in real-world mobile environments. It is fully compatible with both Android smartphones and wearables, facilitating seamless integration and continuous monitoring. 

\begin{figure}[t]
  \centering
  %\vspace{-10pt}
  \includegraphics[width=\linewidth]{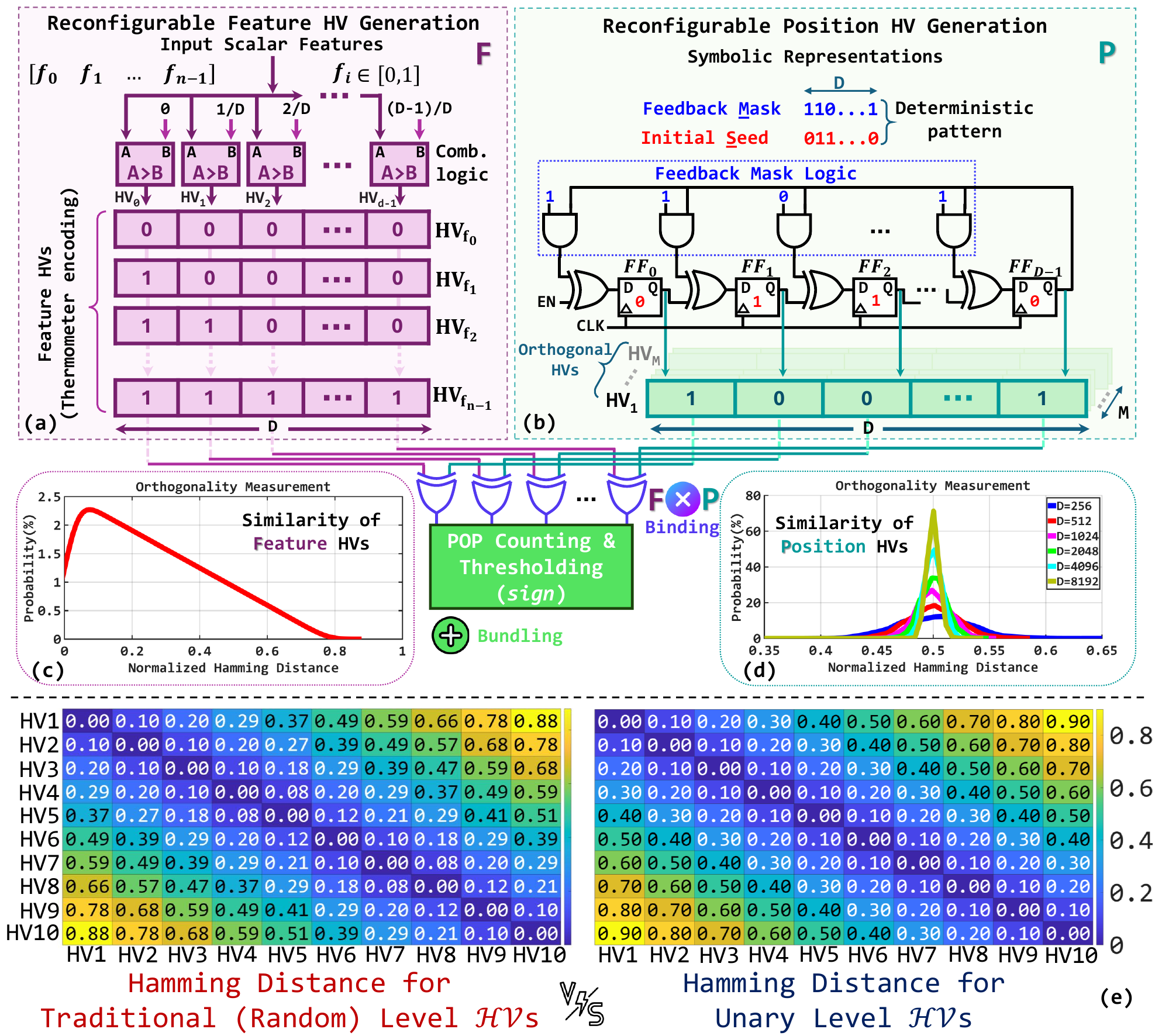}
  \vspace{-1.5em}
  
  \caption{Hardware design strategy for $\mathcal{HV}$ generation. (a) Feature $\mathcal{HV}$s based on \textit{thermometer (unary) encoding}, (b) Position $\mathcal{HV}$s based on a pseudo-LFSR structure and MISR, (c) Hamming distance of feature $\mathcal{HV}$s (\textcolor{Magenta4}{$\mathbf{F}$}), (d) Hamming distance of position $\mathcal{HV}$s (\textcolor{Turquoise4}{$\mathbf{P}$}), \black{and (e) Hamming distance of level $\mathcal{HV}$s using traditional (random) and unary encoding}.
  }  
  \label{HW_design}
  \vspace{-1.5em}
\end{figure}

\subsection{Proposed Hardware %Low-Level 
Design}
\label{HW_design_section}
In addition to the high-level, mobile platform %Android-side monitoring 
described above, this subsection presents a hardware design %circuit-level hardware logic 
that realizes the proposed HDC pipeline. We implement the~\Design~framework on a PYNQ-Z2 FPGA-SoC board equipped with a dual-core ARM Cortex-A9 processor to evaluate % considering the 
binary classification. This heterogeneous platform performs signal preprocessing on the ARM processor, while $\mathcal{HV}$ generation, encoding, and inference are implemented as custom digital logic in the FPGA.
For feature $\mathcal{HV}$ ($F$) encoding, we employed a hardware-efficient, on-the-fly $\mathcal{HV}$ generation method rather than reading $\mathcal{HV}$s from memory. This dynamic approach utilizes \textit{thermometer} (also called \textit{unary}) coding~\cite{Najafi_TVLSI2018_Unary}, where the number of %consecutive 
`1's in the generated $\mathcal{HV}$ increase proportionally with the feature value.  %is proportional to the feature values for the corresponding feature $\mathcal{HV}$, $F$. %(the number of `1's increases stepwise towards higher feature values). 
Each normalized feature $f_i$ in the $[0,1]$ interval is compared against the set $\{0,\frac{1}{D},\frac{2}{D},...,\frac{D-1}{D}\}$ using combinational logic, producing a complete $\mathcal{HV}$ within a single clock cycle. 
For $n$ input features ($n\ll D$), all feature $\mathcal{HV}$s are generated in $n$ clock cycles. 
%For feature values of length $n$, where $n\ll D$, all the feature $\mathcal{HV}$s are generated in $n$ clock cycles. 
Fig.~\ref{HW_design}(a) shows the feature $\mathcal{HV}$ generator module. Using unary structures for vectors offers two main advantages: (i) simplified hardware implementation with reduced switching activity, since groups of \texttt{1}s and \texttt{0}s occur together in the logic, and (ii) a natural correlation property, where closer numerical values lead to more similar $\mathcal{HV}$s, while distant values produce more distinct representations; this is important for feature numerical values.
%On the other hand, the 

\black{In the proposed \Design{} framework, all predictive input parameters--including standard physiological measurements (HR, SPO\textsubscript{2}) as well as temporal and event-based variables--are modeled as continuous variables. Because the dataset relies entirely on continuous input features, the hardware design does not require separate processing pathways for nominal categorical variables. Instead, the entire feature array is processed uniformly using the thermometer (unary) encoding scheme. By treating temporal and event progression as continuous scalar values, the unary encoder naturally preserves their physiological and ordinal correlations; physiological states that are numerically or temporally closer are mapped %together intrinsically map 
to $\mathcal{HV}$s with higher Hamming similarity, accurately reflecting the continuous progression of high-altitude exposure and acclimatization.}

Position $\mathcal{HV}$s ($P$) are generated using a pseudo-LFSR structure combined with %based on adopting a similar approach to the LFSR structure together with 
a Multiple Input Shift Register (MISR)~\cite{Roodsari_OTFGEnc_DATE24,Damiani_MISR_TCAD1990} as depicted in Fig.~\ref{HW_design}(b). In this structure, %a $D$ number of 
$D$ flip-flops (\texttt{FF}s) are connected in sequence, % are utilized, 
where the output of the last \texttt{FF} is fed back %fans out 
to the input of all preceding stages %\texttt{FF}s 
through a \textit{feedback mask logic}. When the %corresponding %feedback logic
mask bit is `1', {the output of the last \texttt{FF} ($FF_{D-1}$) is \texttt{XOR}ed with the output of the current state ($FF_1,FF_2,...$) before being passed to the next stage. %and feeds as an input to the next \texttt{FF}; 
When the mask bit is `0', the feedback is disabled, and the output of the current state is forwarded to the next state.} To ensure producing high--quality, uncorrelated $\mathcal{HV}$s (orthogonal $\mathcal{HV}s$ with an average Hamming distance of 0.5), we adopted a deterministic initialization strategy for \texttt{FF}s, using \textit{initial seeds} and feedback mask patterns derived
%We generated these two $D$-dimensional pattern streams drawn 
from standard Sobol sequences generated in \texttt{MATLAB}. With this approach, each position $\mathcal{HV}$ can be produced in a single clock cycle, requiring $M$ clock cycles to generate $M$ distinct $\mathcal{HV}$s. By replicating this structure with different seeds and feedback masks (e.g., using different random mask patterns like Hadamard with the aid of reconfigurable architecture), multiple $\mathcal{HV}$s can be generated in parallel within a single cycle. This is crucial, especially for longer vectors and increased numbers of samples to be considered for their positions. 

We validated the orthogonality of the generated $\mathcal{HV}$s by measuring Hamming distance across both feature and position $\mathcal{HV}$s, as illustrated in Figs.~\ref{HW_design}(c) and (d).
%We further measured the Hamming distance between distinct output $\mathcal{HV}$s for both structures. Figs.~\ref{HW_design}(c) and (d) represent the similarity of feature and position $\mathcal{HV}$s, respectively.
As can be seen, feature $\mathcal{HV}$s show peaks near zero,  reflecting similarity among closely related features, while position  $\mathcal{HV}$s cluster around 0.5 (strong orthogonality), with the peak narrowing as dimensionality increases. 
\black{Finally, Fig.~\ref{HW_design}(e) demonstrates the Hamming distance comparison between level $\mathcal{HV}$s generated using the traditional approach (through random bit flipping) and thermometer (unary) encoding approach. In both approaches, adjacent $\mathcal{HV}$s representing closer numerical values have near-zero Hamming distance, indicating high similarity, whereas $\mathcal{HV}$s representing distant numerical values approach a Hamming distance near 1, indicating low similarity.
%corresponding to nearby features and the far ones exhibit near 1 (low similarity) due to spaced feature values.
}

\begin{figure}[t]
  \centering
  \includegraphics[width=\linewidth]{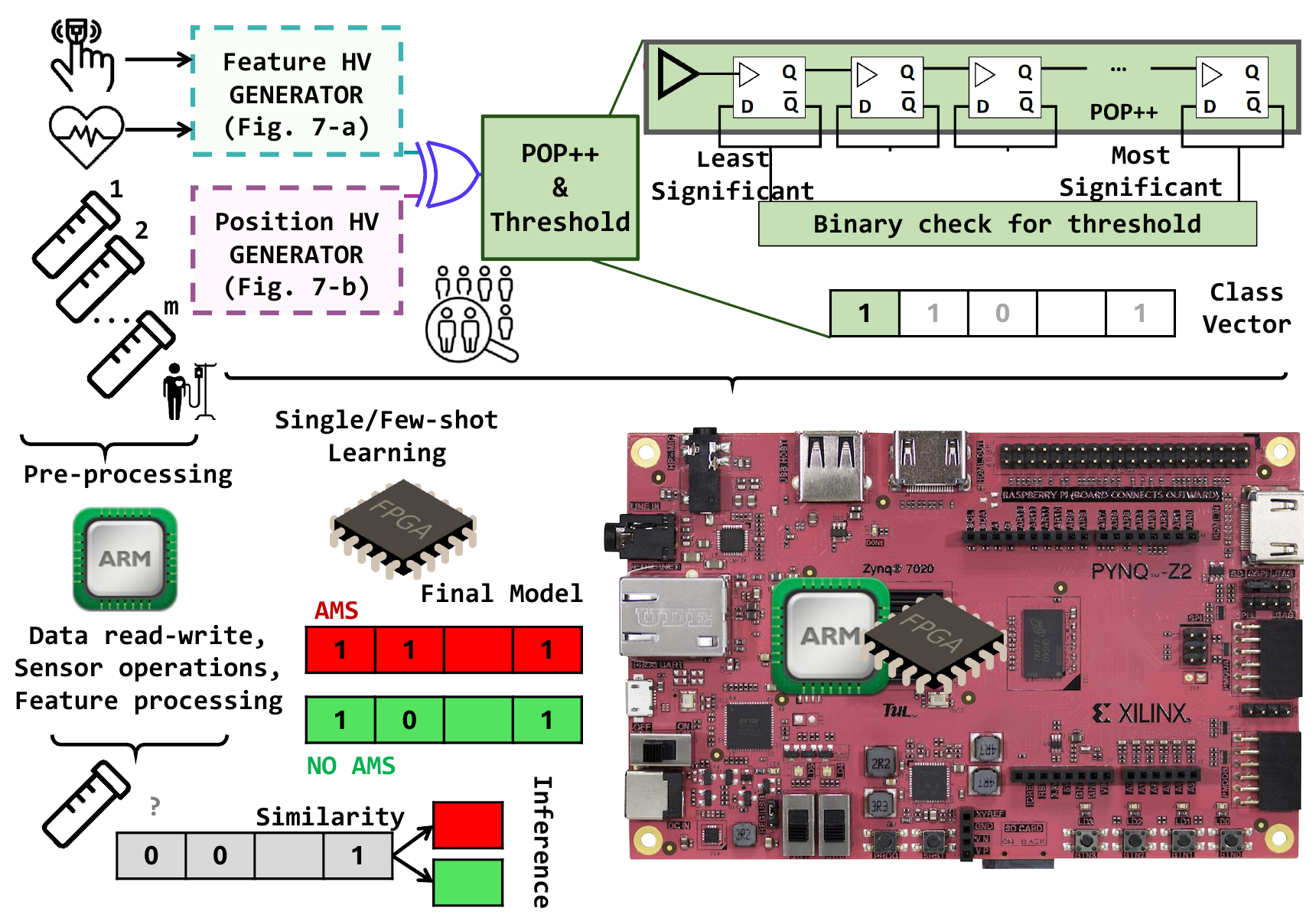}
  \vspace{-1.5em}
  \color{black}
  \caption{Binary computing implementation of~\Design~on PYNQ-Z2 (Zynq-7020). The ARM Cortex-A9 executes data read/write, sensor operations, and feature preprocessing. The FPGA fabric implements on-chip feature/position $\mathcal{HV}$ generation, binding/bundling with accumulative single/few-shot learning to form class $\mathcal{HV}$s, a pipelined popcount–threshold (POP++) stage, and similarity search, yielding the final AMS or No-AMS decision.}  
  \label{lowlevel}
  \vspace{-1.5em}
\end{figure}

Fig.~\ref{lowlevel} details the binary computing implementation on the PYNQ-Z2 (Zynq-7020) FPGA board. The heterogeneous SoC partitions workload between the ARM Cortex-A9 and the FPGA fabric, with the ARM core handling data I/O, interfacing, and feature preprocessing. 
\black{To prepare the physiological signals for hardware-based HDC encoding, the data pass through a lightweight, two-step preprocessing pipeline executed entirely on the ARM processor. First, to account for the different scales and variances of the input signals, each incoming feature $f_i$ is standardized using $z$-score normalization: $z_i = (f_i - \mu) \times (\frac{1}{\sigma})$. For real-time streaming, the mean ($\mu$) and standard deviation ($\sigma$) are not computed on the fly; instead, they are pre-calculated offline during the training phase and stored as static constants. This reduces $z$-score normalization at inference time to one scalar subtraction and one scalar multiplication per feature. 
Following standardization, the $z_i$ values undergo a subsequent clipping and linear min-max mapping to bound them strictly within the $[0, 1]$ interval. This second step is %mathematically
required as the FPGA-based thermometer encoder expects a normalized probability or magnitude bounded between $0$ and $1$ to determine %properly assign 
the ratio of `1's in the generated $\mathcal{HV}$. 
Therefore, the full preprocessing sequence, $z$-score standardization followed by clipping and min-max bounding, is performed on the ARM processor before the data are transmitted to the FPGA. As a result, this preprocessing does not introduce additional FPGA logic-resource overhead.
%there are no hardware costs or overhead on the FPGA logic resources as the
%Because this 
%entire sequence, $z$-score standardization followed by min-max bounding, is handled in the ARM processor prior to hardware transmission. 
In the current hardware pipeline, event-time information is treated as ordered contextual stage information rather than as an arbitrary nominal category. This allows the selected inputs to follow the same scalar preprocessing path, including normalization, bounding, and unary encoding. If future datasets include purely nominal categorical variables without ordinal, temporal, or protocol-stage meaning, such categorical variables can instead be encoded separately using identity $\mathcal{HV}$s and then bound with the corresponding feature-value $\mathcal{HV}$s.}

The FPGA design then implements the end-to-end HDC pipeline: (i) on-chip generation of feature and position $\mathcal{HV}$s, (ii) binding/bundling with accumulative (single/few-shot) learning to form class $\mathcal{HV}$s, and (iii) a pipelined population-count–and-threshold unit (``popcount'') followed by a similarity search against stored class vectors. During inference, each sample follows the same encoding path to produce a query $\mathcal{HV}$; the FPGA then computes the similarity and outputs the binary decision (``\texttt{AMS}'' or ``\texttt{No AMS}''), optionally driving an on-board indicator. %\red{This division keeps the critical path in hardware for low latency and energy, while preserving software flexibility on the ARM. [Better to ignore this sentence as we don't say anything for energy!]}

\section{Evaluation and Results}
\label{evaluation}
In this section, we present a comprehensive evaluation of the proposed \Design~framework. We begin by comparing the performance of conventional ML models with HDC-based methods for both binary and multiclass tasks. We then benchmark our method on external AMS-related datasets and evaluate its deployment efficiency across embedded, FPGA, and mobile-wearable platforms. %with other datasets. 
\textcolor{black}{For model evaluation, the dataset was divided into training and testing sets using an $80$/$20$ sample-wise split: $80$\% of the samples were used for training, and the remaining $20$\% were held out for testing. The same split was applied consistently across all evaluated models, including SVM, multi-layer perceptron (MLP), and the proposed \Design~variants, for both binary and multiclass classification.} Finally, we assess the efficiency of \Design~across multiple hardware platforms, including ARM-based embedded devices, FPGA prototypes, and mobile–wearable systems.

%\subsection{Classical vs. HDC model performance}
%Table~\ref{tab:performance_comparison} compares the performance of different methods for both binary and multiclass classification tasks. The evaluated models include conventional methods of SVM and three HDC encoding strategies: pseudo-random, quasi-random (Sobol), and Hadamard. For binary classification, the Sobol- and Hadamard-based HDC approaches achieved the acclaimed results, followed closely by the pseudo-random approach. In contrast, the conventional ML model lagged significantly. While trying the floating point and the 16-bit representation for SVM, the results are not different in binary calassiication; still not better than any HDC model except the D=128 pseudo-random. Sobol is good at most of the D values, especially for the shorter vector sizes; it is a good alternative in the binary classification case. When it comes to the multi-class, Sobol values fluctuate, but the Hadamard improves after D=1000; so for the longer D size, Hadamar can be an alternative for the multi-class classifier.

\subsection{Classical vs. HDC Model Performance}
\textcolor{black}{Table~\ref{tab:performance_comparison} compares binary and multiclass performance across conventional SVM, MLP, and several HDC encoding approaches. The HDC methods include pseudo-random (\textit{P.-random}), quasi-random (\textit{Sobol}), Hadamard (symbolic) with pseudo-random (level) $\mathcal{HV}$s (\textit{Hadamard} + \textit{P.-random}), Hadamard (symbolic) with Unary (level) $\mathcal{HV}$s (\textit{Hadamard} + \textit{Unary}), and Sobol (symbolic) with threshold-based~\cite{aygun2024sobol} Unary (level) $\mathcal{HV}$s (\textit{Sobol} + \textit{Unary}). For binary classification, the 16-bit fixed-point SVM baseline achieves $0.72$ accuracy, $0.69$ precision, $0.87$ recall, $0.76$ F1-score, and $0.63$ AUC. The MLP baseline, implemented with two hidden layers of 64 and 32 neurons, ReLU activation, Adam optimizer, learning rate = 0.001, and 20 iterations, achieves $0.72$ accuracy, $0.73$ precision, $0.69$ recall, $0.71$ F1-score, and $0.87$ AUC. This indicates strong ranking capability but lower recall and F1-score than several HDC configurations. 
The original HDC encodings generally match or outperform the conventional baselines in binary accuracy and F1-score. Pseudo-random encoding reaches $0.75$ accuracy and $0.78$ F1-score for \(D \ge 512\). Sobol encoding reaches $0.75$ accuracy, $0.72$ precision, $0.88$ recall, and $0.78$ F1-score for \(D \ge 512\). Hadamard encoding also provides stable binary performance across dimensions with $0.75$ accuracy, $0.70$ precision, $0.88$ recall, and $0.78$ F1-score.}

\black{The unary (level $\mathcal{HV}$) configurations further improve binary AMS detection. Hadamard + Unary reaches $0.81$ accuracy at \(D=128\) and improves to $0.84$ accuracy, $0.82$ precision, $0.88$ recall, and $0.85$ F1-score for \(D \ge 256\). Sobol with threshold-based Unary level $\mathcal{HV}$s provides the strongest binary accuracy/F1-score trade-off. In particular, Sobol with threshold (method used to generate symbolic vectors using $0 \leq th \leq 1$, rather than $th$=0.5 fixed value~\cite{aygun2024sobol}), \textit{Sobol w/} \(th=0.65\) + \textit{Unary} at \(D=256\) size achieves the best binary result with $0.91$ accuracy, $1.00$ precision, $0.81$ recall, $0.90$ F1-score, and $0.79$ AUC. Sobol w/ \(th=0.35\) + Unary also remains strong, reaching $0.88$ accuracy at both \(D=128\) and \(D=256\), with AUC up to $0.82$. These results show that threshold-tuned Sobol + Unary encoding can improve binary AMS alert performance while maintaining high precision.}

\black{For multiclass classification, the 16-bit fixed-point baseline SVM achieves $0.67$ accuracy, $0.66$ precision, $0.66$ recall, $0.65$ F1-score, and $0.71$ AUC. The MLP baseline achieves $0.75$ accuracy, $0.81$ precision, $0.74$ recall, $0.66$ F1-score, and $0.91$ AUC, indicating strong probability-based class separability but limited F1-score improvement. The original HDC methods improve multiclass performance over SVM, with pseudo-random and Hadamard encodings reaching up to $0.70$ accuracy, $0.69$ precision, $0.70$ recall, $0.69$ F1-score, and $0.73$ AUC. Sobol encoding achieves slightly lower accuracy of $0.68$ but provides the highest AUC among the original HDC encodings, reaching $0.74$, indicating stronger ROC-style class separability. The Unary-level configurations also remain competitive in the multiclass setting. Hadamard + Unary reaches $0.73$ accuracy, $0.72$ precision, $0.73$ recall, $0.72$ F1-score, with AUC in the range of $0.71$--$0.72$. Among the Sobol threshold-based Unary settings, Sobol w/ \(th=0.75\) + Unary achieves the strongest multiclass result at compact dimensions, with $0.76$ accuracy, $0.78$ precision, $0.76$ recall, $0.75$ F1-score, and $0.75$ AUC. Overall, these results verify that unary level $\mathcal{HV}$s are effective on the AMS dataset, improving binary AMS detection while preserving competitive multiclass severity classification.}

%===============================================
\begin{table}[t]
    \centering
    \color{black}
    \caption{Performance Comparison of Classification Approaches. The \Design{} Methods are Based on Single Pass Iteration %Attains High Accuracy in a Single Iteration.
    }
    \vspace{-0.5em}
    %\scriptsize
    \setlength{\tabcolsep}{2.8pt}
    \resizebox{\columnwidth}{!}{%
    \begin{tabular}{|l|c|ccccc|ccccc|}
    \hline
    \multirow{2}{*}{\textbf{Classification Approach}} & \multirow{2}{*}{\textbf{D}} 
    & \multicolumn{5}{c|}{\textbf{Binary}} 
    & \multicolumn{5}{c|}{\textbf{Multiclass}} \\ \cline{3-12}
    & & \textbf{Acc} & \textbf{Pre} & \textbf{Rec} & \textbf{F1} & \textbf{AUC}
    & \textbf{Acc} & \textbf{Pre} & \textbf{Rec} & \textbf{F1} & \textbf{AUC} \\ \hline\hline

    %\textbf{SVM (floating point)} & -- & 0.72 & 0.67 & 0.88 & 0.76 & 0.63 & 0.67& 0.67 & 0.67 & 0.66 & 0.71 \\ \hline
    \textbf{SVM (16-bit fixed point)} & -- & 0.72 & 0.69 & 0.87 & 0.76 & 0.63 & 0.67 & 0.66 & 0.66 & 0.65 & 0.71 \\ \hline
    \textbf{MLP} & -- & 0.72 & 0.73 & 0.69 & 0.71 & 0.87 & 0.75 & 0.81 & 0.74 & 0.66 & 0.91 \\ \hline
    
    \multirow{6}{*}{\begin{tabular}[l]{@{}l@{}} \Design\\ \textbf{(P.-random + P.-random)} \end{tabular}}   
    & 128   & 0.72 & 0.67 & 0.88 & 0.76 & 0.63 & 0.67 & 0.66 & 0.66 & 0.65 & 0.71 \\ %\cline{2-12}
    & 256   & 0.72 & 0.67 & 0.88 & 0.76 & 0.63 & 0.70 & 0.69 & 0.70 & 0.69 & 0.73 \\ %\cline{2-12}
    & 512   & 0.75 & 0.70 & 0.88 & 0.78 & 0.66 & 0.70 & 0.69 & 0.70 & 0.69 & 0.73 \\ %\cline{2-12}
    & 1024  & 0.75 & 0.70 & 0.88 & 0.78 & 0.66 & 0.70 & 0.69 & 0.70 & 0.69 & 0.73 \\ %\cline{2-12}
    & 2048  & 0.75 & 0.70 & 0.88 & 0.78 & 0.66 & 0.70 & 0.69 & 0.70 & 0.69 & 0.73 \\ %\cline{2-12}
    & 10000 & 0.75 & 0.70 & 0.88 & 0.78 & 0.66 & 0.70 & 0.69 & 0.70 & 0.69 & 0.73 \\ \hline

    \multirow{6}{*}{\begin{tabular}[l]{@{}l@{}} \Design\\ \textbf{(Sobol + P.random)} \end{tabular}}    
    & 128   & 0.69 & 0.67 & 0.75 & 0.71 & 0.65 & 0.65 & 0.64 & 0.65 & 0.64 & 0.74 \\ %\cline{2-12}
    & 256   & 0.75 & 0.72 & 0.81 & 0.76 & 0.65 & 0.67 & 0.66 & 0.66 & 0.65 & 0.74 \\ %\cline{2-12}
    & 512   & 0.75 & 0.72 & 0.88 & 0.78 & 0.65 & 0.68 & 0.67 & 0.68 & 0.67 & 0.74 \\ %\cline{2-12}
    & 1024  & 0.75 & 0.72 & 0.88 & 0.78 & 0.65 & 0.68 & 0.67 & 0.68 & 0.67 & 0.74 \\ %\cline{2-12}
    & 2048  & 0.75 & 0.72 & 0.88 & 0.78 & 0.65 & 0.68 & 0.67 & 0.68 & 0.67 & 0.74 \\ %\cline{2-12}
    & 10000 & 0.75 & 0.72 & 0.88 & 0.78 & 0.65 & 0.68 & 0.67 & 0.68 & 0.67 & 0.74 \\ \hline

    \multirow{6}{*}{\begin{tabular}[l]{@{}l@{}} \Design\\ \textbf{(Hadamard + P.random)} \end{tabular}} 
    & 128   & 0.75 & 0.70 & 0.88 & 0.78 & 0.61 & 0.67 & 0.66 & 0.66 & 0.65 & 0.72 \\ %\cline{2-12}
    & 256   & 0.75 & 0.70 & 0.88 & 0.78 & 0.63 & 0.67 & 0.66 & 0.66 & 0.65 & 0.72 \\ %\cline{2-12}
    & 512   & 0.75 & 0.70 & 0.88 & 0.78 & 0.63 & 0.70 & 0.69 & 0.70 & 0.69 & 0.73 \\ %\cline{2-12}
    & 1024  & 0.75 & 0.70 & 0.88 & 0.78 & 0.63 & 0.70 & 0.69 & 0.70 & 0.69 & 0.73 \\ %\cline{2-12}
    & 2048  & 0.75 & 0.70 & 0.88 & 0.78 & 0.63 & 0.70 & 0.69 & 0.70 & 0.69 & 0.73 \\ %\cline{2-12}
    & 10000 & 0.75 & 0.70 & 0.88 & 0.78 & 0.63 & 0.70 & 0.69 & 0.70 & 0.69 & 0.73 \\ \hline

    \multirow{6}{*}{\begin{tabular}[l]{@{}l@{}} \Design\\ \textbf{(Hadamard + Unary)} \end{tabular}} 
    & 128   & 0.81 & 0.78 & 0.88 & 0.82 & 0.73 & 0.70 & 0.68 & 0.69 & 0.68 & 0.72 \\ %\cline{2-12}
    & 256   & 0.84 & 0.82 & 0.88 & 0.85 & 0.72 & 0.73 & 0.72 & 0.73 & 0.72 & 0.71 \\ %\cline{2-12}
    & 512   & 0.84 & 0.82 & 0.88 & 0.85 & 0.71 & 0.73 & 0.72 & 0.73 & 0.72 & 0.71 \\ %\cline{2-12}
    & 1024  & 0.84 & 0.82 & 0.88 & 0.85 & 0.71 & 0.73 & 0.72 & 0.73 & 0.72 & 0.71 \\ %\cline{2-12}
    & 2048  & 0.84 & 0.82 & 0.88 & 0.85 & 0.70 & 0.73 & 0.72 & 0.73 & 0.72 & 0.72 \\ %\cline{2-12}
    & 10000 & 0.84 & 0.82 & 0.88 & 0.85 & 0.70 & 0.73 & 0.72 & 0.73 & 0.72 & 0.71 \\ \hline
    
    \multirow{2}{*}{\begin{tabular}[l]{@{}l@{}} \Design \\ \textbf{(Sobol w/ th=0.35 + Unary)} \end{tabular}} 
    & 128 & 0.88 & 1.00 & 0.75 & 0.86 & 0.82 & 0.71 & 0.72 & 0.71 & 0.70 & 0.73 \\ %\cline{2-12}
    & 256 & 0.88 & 0.93 & 0.81 & 0.87 & 0.75 & 0.71 & 0.72 & 0.70 & 0.70 & 0.70 \\ \hline
    
    \multirow{2}{*}{\begin{tabular}[l]{@{}l@{}} \Design\\ \textbf{(Sobol w/ th=0.45 + Unary)} \end{tabular}} 
    & 128 & 0.81 & 0.92 & 0.69 & 0.79 & 0.75 & 0.71 & 0.70 & 0.71 & 0.70 & 0.75 \\ %\cline{2-12}
    & 256 & 0.88 & 0.93 & 0.81 & 0.87 & 0.70 & 0.71 & 0.70 & 0.71 & 0.70 & 0.75 \\ \hline

    \multirow{2}{*}{\begin{tabular}[l]{@{}l@{}} \Design\\ \textbf{(Sobol w/ th=0.65 + Unary)} \end{tabular}} 
    & 128 & 0.78 & 0.91 & 0.62 & 0.74 & 0.78 & 0.75 & 0.75 & 0.74 & 0.74 & 0.73 \\ %\cline{2-12}
    & 256 & 0.91 & 1.00 & 0.81 & 0.90 & 0.79 & 0.75 & 0.75 & 0.74 & 0.74 & 0.70 \\ \hline

    \multirow{2}{*}{\begin{tabular}[l]{@{}l@{}} \Design\\ \textbf{(Sobol w/ th=0.75 + Unary)} \end{tabular}} 
    & 128 & 0.81 & 0.92 & 0.69 & 0.79 & 0.82 & 0.76 & 0.78 & 0.76 & 0.75 & 0.75 \\ %\cline{2-12}
    & 256 & 0.81 & 0.92 & 0.69 & 0.79 & 0.79 & 0.76 & 0.78 & 0.76 & 0.75 & 0.75 \\ \hline
    \end{tabular}%
    }
    \vspace{-0.75em}
    \justify{\scriptsize{
    The \textit{threshold (th)} value indicates the biased ratio of `1's in the generated position $\mathcal{HV}$s (P) using the Sobol sequences~\cite{aygun2024sobol}. The pair (P + L; e.g., \textit{Hadamard} + \textit{Unary}) in each \Design{} method indicates the position $\mathcal{HV}$ source (P) and the level $\mathcal{HV}$ source (L).
    }}
    \label{tab:performance_comparison}
    \vspace{-10pt}
\end{table}

\begin{table}[!b]
\centering
\vspace{-1em}
\caption{\Design{} Performance on Different SOTA Datasets during Inference (Binary Classification)}
\vspace{-0.5em}
\label{tab:proposed_sota_inference}
\resizebox{\columnwidth}{!}{
\begin{tabular}{|l|l|c|c|c|c|c|}
\hline
\textbf{ Dataset} & \textbf{Task} & \textbf{D} & \textbf{P-ACC} & \textbf{S-ACC} & \textbf{H-ACC} & \textbf{Time (ms)} \\
\hline
Boos et al.~\cite{boos2018relationship} & Multi & 2000 & 0.76 & 0.76 & 0.76 & 40.42 \\
\hline
Berger et al.~\cite{berger2023prevalence}& Multi & 2000 &  0.84& 0.85 & 0.85 & 44.65  \\
\hline
\end{tabular}
}
\justify{\scriptsize{P: Pseudo-random, S: Sobol, H: Hadamard }}
%\vspace{-15pt}
\end{table}

\subsection{Inference Performance on State-of-the-Art Datasets}

Table~\ref{tab:proposed_sota_inference} exhibits the proposed method's performance within the prior SOTA datasets. Boos et al.\cite{boos2018relationship} present their dataset of $80$ participants. When evaluated with our method, the pseudo-randomness-based accuracy (P-ACC), Sobol-based accuracy (S-ACC), and Hadamard-based accuracy (H-ACC) each reached $0.76$. Berger et al.~\cite{berger2023prevalence} investigated a larger cohort of $1,370$ mountaineers across multiple altitudes ($2,850$~m: $212$ subjects; $3,050$~m: $98$; $3,650$~m: $629$; $4,559$~m: $431$). 
Applying our method to this dataset yielded the highest accuracy among prior studies, with P-ACC of $0.84$, S-ACC of $0.85$, and H-ACC of $0.85$.

\begin{figure}[!t]
%\vspace{-1em}
    \centering
    \includegraphics[width=0.85\linewidth]{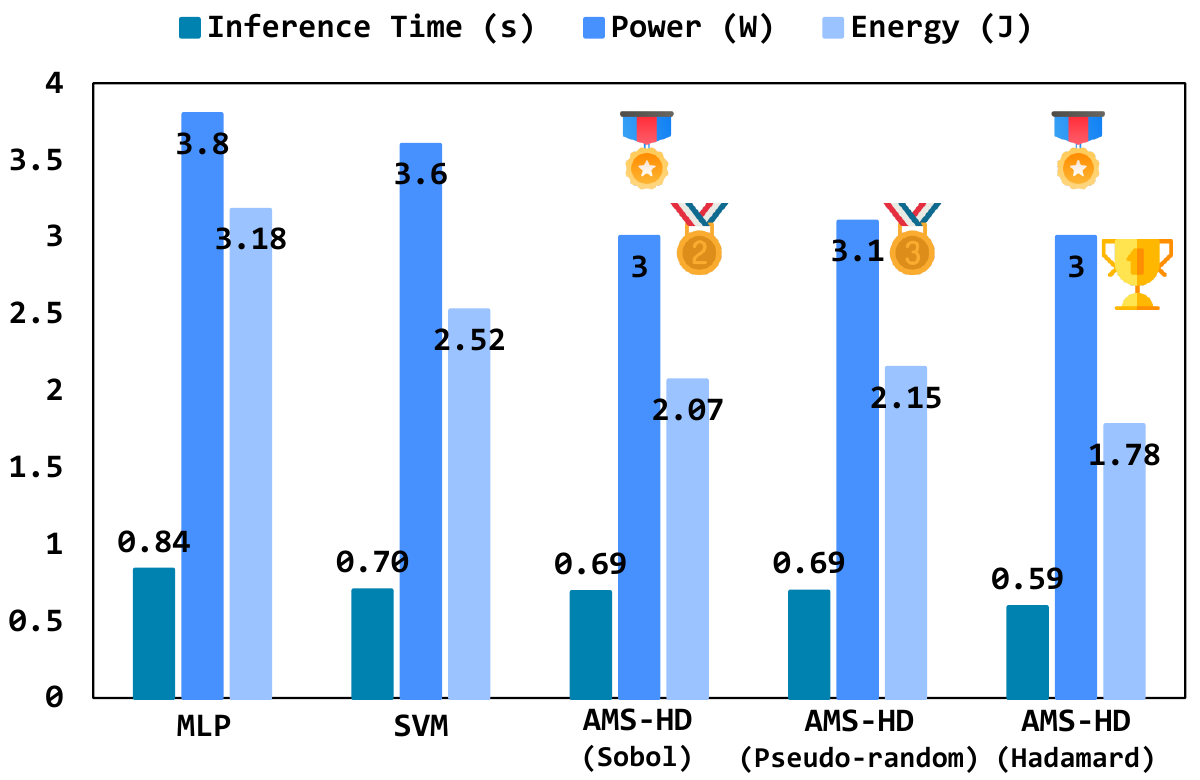}
    \vspace{-0.5em}
    \caption{Comparison of Power Consumption, Inference Time, and Energy Efficiency of different ML algorithms and~\Design~on an ARM Processor.}
    \label{fig:arm_power}
   \vspace{-0.75em}
\end{figure}

\black{In addition to the external dataset evaluations summarized in Table~\ref{tab:proposed_sota_inference}, we further evaluated new-user generalization using an external test-only setting. Specifically, we used the AMS-related dataset from Boos et al.~\cite{boos2018relationship}, which includes similar wearable-accessible physiological variables and Lake Louise score-based AMS annotations. The \Design{} model was trained on the original Pham et al. dataset~\cite{pham2022inflammatory} and then evaluated directly on the external dataset without retraining. For this external validation, we used the Sobol+Unary configuration with threshold=0.75 and $D$=128, which provides a strong overall trade-off in Table~\ref{tab:performance_comparison} for both binary and multiclass settings. This test-only evaluation achieved $82\%$ binary classification accuracy and $79\%$ multiclass classification accuracy. Since the external dataset contains different subjects and was not used during training, these results provide additional evidence that the learned \Design{} representation can generalize beyond the original sample-level split. At the same time, we interpret this analysis as supportive external validation rather than a replacement for a larger prospective subject-wise clinical study, which remains an important future direction.}

\subsection{Embedded Platform Performance}

The execution of conventional ML baselines (MLP: Multi-Layer Perceptron and SVM: Support Vector Machine), along with Sobol, Pseudo-random, and Hadamard-based~\Design, was evaluated on an ARM embedded platform (700 MHz, 32-bit single-core processor, 512 MB RAM). As shown in Fig.~\ref{fig:arm_power}, MLP (with two hidden layers, 300 and 50 neurons at each) is the least efficient, requiring $0.836$ seconds of inference time and consuming $3.8$ W of power with total energy usage of $3.177$ J. SVM improves latency to $0.701$ seconds with a lower energy consumption of $2.523$ J, though its power draw remains relatively high at $3.6$ W. Among~\Design~variants, Sobol achieves an inference time of $0.689$ seconds and the lowest energy usage of $2.067$ J, making it highly efficient. The Pseudo-random method yields comparable latency ($0.693$ seconds) %inference 
with $3.1$ W power consumption and $2.148$ J energy usage, placing it between SVM and Sobol in efficiency. %Notably, 
The Hadamard variant demonstrates the best overall performance, with the shortest inference time of $0.592$ seconds, moderate power consumption of $3.0$ W, and the lowest total energy usage of $1.776$ J. These findings highlight Hadamard encoding as the most effective configuration for ARM-based embedded systems, providing the best trade-off among power, latency, and energy efficiency compared to both conventional ML models and other~\Design~variants.

%\color{teal}
\subsection{FPGA-based Execution}

\begin{table}[!t]
\centering
%\setlength{\tabcolsep}{2pt} 
%\caption{Hardware implementation cost...}
\caption{Hardware Resource Utilization and Power Consumption of Classifiers on FPGA Implemented with 100MHz Frequency}
%\vspace{-0.5em}
\resizebox{\columnwidth}{!}{
\begin{tabular}{|c|c|c|c|c|c|c|c|}
\hline
\textbf{Classifier} & \textbf{Configuration} & \textbf{LUT} & \textbf{FF} & \textbf{BRAM} &  \textbf{DSP} & \begin{tabular}[c]{@{}c@{}} \textbf{Clock}\\ \textbf{Cycles} \end{tabular} & \begin{tabular}[c]{@{}c@{}} \textbf{Total}\\ \textbf{Power (W)} \end{tabular} \\
\hline\hline
\Design & \begin{tabular}[c]{@{}c@{}c@{}} Binary Computing\\ D=256\\ \textbf{Acc. = $91$}\% \end{tabular} & 1257 & 1851 & 1 & 0 & 16 & 1.6  \\
\hline
SVM & \begin{tabular}[c]{@{}c@{}} 16-bit precision\\ \textbf{Acc. = $72$}\% \end{tabular} & 379 & 481 & 0 & 1 & 42 & 2.5 \\
\hline
MLP & \begin{tabular}[c]{@{}c@{}} 16-bit precision\\ \textbf{Acc. = $72$}\% \end{tabular} & 9158 & 10993 & 3 & 207 & 16 & 6.2 \\
\hline
\hline
\begin{tabular}[c]{@{}c@{}c@{}} \Design \\ ISO \\ Accuracy \end{tabular} & \begin{tabular}[c]{@{}c@{}c@{}} Binary Computing\\ D=128\\ \textbf{Acc. = $78$}\% \end{tabular} & 631 & 923 & 1 & 0 & 14 & 0.34  \\ 
\hline
\end{tabular}
}
\label{fpga_report}
\vspace{-0.5em}
\end{table}

We described \Design, SVM, and MLP classifiers in VHDL, and synthesized and implemented them on the FPGA using the Xilinx Vivado Design Suite as reported in Table~\ref{fpga_report}. %To ensure a fair comparison, 
The \Design~classifier (with Sobol) was designed with $\mathcal{HV}$ dimensionalities of $D$=256 and $D$=128 (for iso-accuracy comparison), while the SVM and MLP (with 300- and 50-neuron hidden layer sizes) classifiers were implemented using 16-bit fixed-point precision.
%The proposed \Design~design surpasses the MLP classifier by 46\%, 34\%, and 58\% reductions in terms of Look Up Tables (LUT), \texttt{FF}s and power consumption, respectively. Although, the SVM classifier provides lower LUT and \texttt{FF} utilization compared to the \Design, their overall power consumption are close to each other. Furthermore, considering the performance comparison results discussed in Table~\ref{tab:performance_comparison}, the \Design~design provides more accurate results. 
In terms of resource utilization, the proposed \Design~design shows significant advantages over the MLP, achieving reductions of 7.3$\times$ %46\% 
in the number of Look-Up Tables (LUTs), 5.8$\times$ %34\% 
in \texttt{FF}s, and 3.9$\times$ %58\% 
in power consumption. 
\black{Furthermore, the \Design{} framework surpasses the SVM by 36\% and 62\% reduction in total power consumption and latency (number of clock cycles), respectively.} 
%Although the SVM classifier utilizes fewer LUTs and FFs compared to \Design, their overall power consumption levels are comparable when $D$=1000. 
For $D$=128, \Design~achieves significantly lower power consumption than the SVM design while maintaining higher (ISO) accuracy. As detailed in Table~\ref{tab:performance_comparison}, our \Design~design delivers significantly higher classification accuracy than both the SVM and MLP approaches in general.

\ignore{
\red{[Hassan: We can keep this paragraph only if we report noise injection results that are really bad for SVM and MLP. With current numbers it's better to not discuss robustness to noise at all!!]
As it can be seen in Table~\ref{tab:performance_comparison}, another key advantage of the \Design~architecture is its inherent robustness to soft errors, such as bit-flips induced by noise. This resilience arises from the distributed encoding of information across high-dimensional $\mathcal{HV}$s, which allows the system to tolerate errors without catastrophic failure. As a result, \Design~degrades gracefully under relatively high bit-flip rates (e.g., 10\%). %(logic-1$\leftrightarrow$ logic-0). 
By contrast, conventional binary architectures such as SVM and MLP are highly sensitive soft-errors, where even a small number of bit-flips %even single-bit errors, 
can significantly distort numerical values and corrupt the final output.}
}

\ignore{
\begin{figure*}[t]
  \centering
  \includegraphics[width=\linewidth]{Figures/fpga_ams.pdf}
  %\vspace{-1.5em}
  \color{black}
  \caption{Overall design setup on the PYNQ-Z2 FPGA board. \black{[Will be updated.]}}  
  \label{fpga_board}
  %\vspace{-1.em}
\end{figure*}
}

\begin{table}[t]
    \centering
    \caption{Mobile Platform Performance Summary ($D$=128) %for SVM, MLP, and \Design
    }
    \vspace{-0.5em}
    \resizebox{\columnwidth}{!}{
    \begin{tabular}{|c|c|c|c|}
        \hline
        \textbf{Model} & \textbf{Battery Drop (\%)} & \textbf{Avg Exec. Time (ms)} & \textbf{Memory (Bytes)} \\
        \hline\hline
        SVM & 4 & 3.66 & 180 \\
        \hline
        MLP & 8 & \textbf{0.49} & 276 \\
        \hline
        \Design & \textbf{1} & 2.50 & \textbf{60} \\
        \hline
    \end{tabular}
    }
    \label{tab:hardware_summary_clean}
\vspace{-0.75em}
\end{table}

\color{black}

%\subsection{Mobile-Wearable Platform} %Android-based Experiments

\subsection{Mobile Platform Performance %Android-based Findings
}

Table~\ref{tab:hardware_summary_clean} summarizes the model performance of SVM, MLP, and HDC on the mobile %Android 
platform in terms of battery usage, %impact, 
execution time, and memory footprint. SVM consumed $4\%$ battery with an average execution time of $3.66ms$ and a memory requirement of $180$~Bytes. Battery drain is monitored across the ten-minute test using Android’s built-in \textit{BatteryManager} application programming interface. MLP achieved the fastest execution at $0.49ms$, but at the cost of higher battery consumption ($8\%$) and memory usage ($276$~Bytes). In contrast, HDC demonstrated the most energy-efficient performance, requiring only a $1\%$ battery drop and the smallest memory footprint ($60$~Bytes) for the deployed model, while maintaining an average execution time of $2.50ms$. These results highlight the trade-offs across models: MLP provides minimal latency, whereas HDC offers superior energy efficiency and compact memory usage, making it the most practical %suitable 
choice for deployment on resource-constrained mobile and wearable devices.

\begin{figure}[!t]
    \centering
    \includegraphics[width=0.73\linewidth]{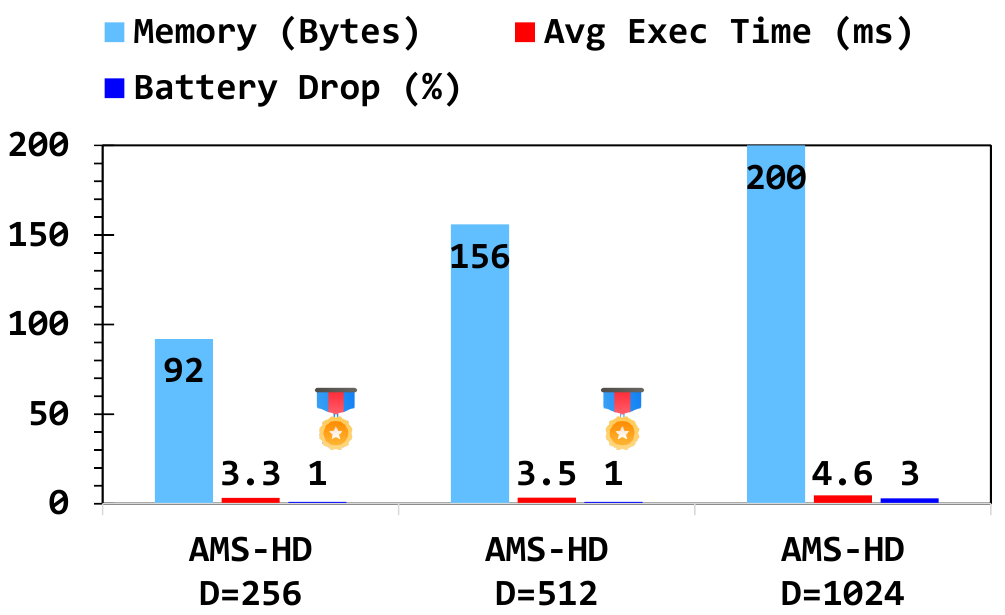}
    \vspace{-0.5em}
    \caption{Comparison of Model Memory Usage, Execution Time, and Battery Efficiency across different HDC model sizes ($D$).}
    \label{fig:hard_hdc}
    \vspace{-0.5em}
\end{figure}

\begin{table}[!t]
\centering
\color{black}
\caption{%Average instantaneous current, accumulated charge (mAh), and energy consumption across models on a mobile platform. Each configuration was benchmarked for 5 minutes and repeated twice; reported values are averaged across runs after accounting for the idle/background baseline reference stabilization.
Comparison of Energy and Charge Consumption on the Mobile Platform
} 
\vspace{-0.5em}
%\footnotesize
%\setlength{\tabcolsep}{4pt}
%\renewcommand{\arraystretch}{1.05}
\resizebox{\columnwidth}{!}{
\begin{tabular}{|l|c|c|c|}
\hline
\textbf{Model} & \textbf{Avg Current (mA)} & \textbf{Charge (mAh)} & \textbf{Energy (mWh)} \\ \hline\hline
\textbf{\Design{} ($D$=128)}  & 61.34  & 5.11  & 19.83 \\ \hline
\textbf{\Design{} ($D$=256)}  & 66.80  & 5.57  & 21.60 \\ \hline
\textbf{\Design{} ($D$=512)}  & 74.90  & 6.24  & 24.22 \\ \hline
\textbf{\Design{} ($D$=1024)} & 81.97  & 6.83  & 26.50 \\ \hline
\textbf{SVM}          & 123.56 & 10.30 & 39.95 \\ \hline
\textbf{MLP}          & 212.14 & 17.68 & 68.59 \\ \hline
\end{tabular}
}
\vspace{-0.75em}
\justify{\scriptsize{
Each configuration was benchmarked for 5 minutes and repeated twice; reported values are averaged across runs after accounting for the idle baseline reference stabilization.
}}
\label{tab:arm_energy}
\vspace{-1em}
\end{table}

\begin{table*}[!t]
\centering
%\vspace{-15pt}
\caption{Hardware Implementation Comparison of HDC Acceleration Frameworks}
\resizebox{\textwidth}{!}{
\begin{tabular}{|c|c|c|c|c|c|c|c|c|c|} 
\hline
\multirow{2}{*}{\textbf{Framework}} & \multirow{2}{*}{\textbf{Target FPGA}} & \multirow{2}{*}{\textbf{Key Contribution / Encoding Method}} & \multicolumn{4}{c|}{\textbf{Resource Utilization}} & \multirow{2}{*}{\begin{tabular}[c]{@{}c@{}c@{}} \textbf{Power}\\ (W) \end{tabular}} & \multirow{2}{*}{\begin{tabular}[c]{@{}c@{}c@{}} \textbf{Freq.}\\ (MHz) \end{tabular}} & \multirow{2}{*}{\textbf{Remarks}} \\ 
\cline{4-7}
 &  &  & \textbf{LUT} & \textbf{FF} & \textbf{BRAM} & \textbf{DSP} &  &  &  \\ 
\hline
E3HDC~\cite{Roodsari_E3HDC_2024} & PYNQ-Z2 & \begin{tabular}[c]{@{}c@{}c@{}} On-the-fly $\mathcal{HV}$\\
  generation to eliminate BRAM \end{tabular} & 12883 & 21759 & 0 & 0 & 0.259 & 120 & \begin{tabular}[c]{@{}c@{}c@{}} D=1K for MNIST dataset;\\ Binary model. \end{tabular} \\ 
\hline
HD2FPGA~\cite{Zhang_HD2FPGA} & Xilinx U280 & \begin{tabular}[c]{@{}c@{}c@{}} Automated tool for
  classification\\ (Random Projection) \end{tabular} & -- & -- & -- & -- & -- & -- & 1.5$\times$ speedup over
  F5-HD. \\ 
\hline
\multirow{3}{*}{F5-HD~\cite{Salamat_F5HD} } & Xilinx Kintex-7 & \begin{tabular}[c]{@{}c@{}c@{}} Template-based framework\\
  with multiple precisions \end{tabular} & 46\% & -- & 47\% & 89\% & 9.8 & -- & Fixed-point model. \\ 
\cline{2-10}
 & Xilinx Kintex-7 & Binary model
   & 82\% & -- & 5\% & 29\% & 13.9 & -- & \begin{tabular}[c]{@{}c@{}c@{}} D=10K;
  Higher logic usage\\ far less BRAM/DSP. \end{tabular} \\ 
\hline
QuantHD~\cite{Imani_QuantHD} & Xilinx Kintex-7 & \begin{tabular}[c]{@{}c@{}c@{}} Quantization framework to\\ reduce model size \end{tabular} & $\approx$16K & $\approx$19K & 7 & -- & -- & 100 & \begin{tabular}[c]{@{}c@{}c@{}} D=1K for ISOLET
  dataset; \\Binary model. \end{tabular} \\ 
\hline
\Design & PYNQ-Z2 & \begin{tabular}[c]{@{}c@{}c@{}} On-the-fly $\mathcal{HV}$ generation\\ for biomedical signals \end{tabular} & 1257 & 1851 & 1 & 0 & 1.6 & 100 & \begin{tabular}[c]{@{}c@{}c@{}} D=256 for AMS dataset; Binary model.\\ High Accuracy with low Dimensionality.\end{tabular} \\
\hline
\end{tabular}
}
\label{tab:hardware_comparison_expanded}
\vspace{-12pt}
\end{table*}

As illustrated in Fig.~\ref{fig:hard_hdc}, increasing the $\mathcal{HV}$ dimensionality ($D$) directly impacts memory usage, % Fig.~\ref{fig:hard_hdc} shows its effect on memory, 
execution time, and battery consumption on the mobile %Android 
platform. When $D$ grows from 256 to 1024, memory usage rises from $92$ Bytes (HDC-256) to $200$ Bytes (HDC-1024). Similarly, the average execution time increases %accordingly, 
from $3.34ms$ at $D=256$ to $4.57ms$ at $D=1024$. Battery consumption remains minimal at $1\%$ for HDC-256 and HDC-512, but increases to $3\%$ once $D=1024$. These results indicate a near-linear scaling trend in memory and execution time, with battery efficiency beginning to degrade as dimensionality exceeds 512.

\black{Table~\ref{tab:arm_energy} reports the average current draw, accumulated charge (mAh), and energy consumption of each model on the mobile platform. Measurements were collected from fixed-duration benchmark runs (5 minutes), during which the models executed continuous inference workloads. The reported values are averaged across repeated runs. To account for baseline mobile power consumption, the phone/application was first kept in a steady idle state for a 5-minute interval before model execution. This averaged idle measurement was then used to evaluate the additional inference workload under the same device and application conditions. The accumulated charge is computed over the runtime and can be linearly scaled for comparison with longer-duration evaluations. The results show that HDC-based models achieve substantially lower energy and charge consumption compared to conventional approaches. For example, HDC with $D=128$ consumes 61.34\,mA, corresponding to 5.11\,mAh and 19.83\,mWh, whereas SVM and MLP consume 123.56\,mA / 10.30\,mAh / 39.95\,mWh and 212.14\,mA / 17.68\,mAh / 68.59\,mWh, respectively. This corresponds to approximately $2\times$ lower energy and charge consumption than SVM and more than $3\times$ lower than MLP. Although energy and charge consumption increase gradually with HDC dimensionality, even the largest configuration, $D=1024$, remains significantly more efficient than the conventional baselines. These results highlight the suitability of HDC for energy-efficient inference on mobile devices.}

%\textcolor{purple}{Table~\ref{tab:arm_energy} reports the average current draw, accumulated charge (mAh), and energy consumption of each model on a mobile platform. Measurements are obtained from fixed-duration benchmark runs (5 minutes), during which models execute continuous inference workloads, and the reported values are averaged over different runs. Th be fairly report the baseline idle phone comnsumtipons the phone was stabizled in the idele for a time... The accumulated charge is computed over the runtime and can be linearly scaled for comparison with longer-duration evaluations. The results show that HDC-based models achieve substantially lower energy and battery usage compared to conventional approaches. For example, HDC ($D=128$) consumes 61.34\,mA, corresponding to 5.11\,mAh and 19.83\,mWh, whereas SVM and MLP consume 123.56\,mA / 10.30\,mAh / 39.95\,mWh and 212.14\,mA / 17.68\,mAh / 68.59\,mWh, respectively. This corresponds to approximately $2\times$ lower energy and charge consumption than SVM and more than $3\times$ lower than MLP. Although energy and charge consumption increase gradually with HDC dimensionality, even the largest configuration ($D=1024$) remains significantly more efficient than the conventional baselines. These results highlight the suitability of HDC for energy-efficient inference on mobile devices.}

\section{Comparison with SOTA Works}
\label{Sec:SOTA}

\subsection{HDC Accelerators Comparison}

\color{black}
Table~\ref{tab:hardware_comparison_expanded} positions the proposed \Design{} framework within the broader landscape of FPGA-based HDC accelerators. A direct comparison on the same edge-focused PYNQ-Z2 platform highlights \Design’s significant resource savings. By leveraging the optimized $D=256$ dimensionality, which preserves high accuracy for the AMS dataset, our design utilizes approximately 10.2$\times$ fewer LUTs and 11.8$\times$ fewer FFs than E3HDC~\cite{Roodsari_E3HDC_2024}, which operates at $D=1000$. This contrast highlights a key trade-off: while E3HDC achieves its novel BRAM-less encoding at the cost of higher logic utilization, \Design{} offers a highly balanced and compact design with a minimal memory footprint of only one BRAM block.
Furthermore, when compared to QuantHD~\cite{Imani_QuantHD}—a quantization-focused framework implemented on the Kintex-7 FPGA—AMS-HD requires significantly fewer hardware resources (1,257 vs. 16,000 LUTs, and 1 vs. 7 BRAMs) while operating at the same 100 MHz frequency.
When compared with high-performance frameworks, differing design philosophies become apparent. F5-HD~\cite{Salamat_F5HD}, implemented on a more powerful Kintex-7 FPGA, demonstrates flexibility by supporting a high-accuracy fixed-point model that heavily utilizes DSPs (89\%), as well as a binary model ($D=10,000$) that shifts the computational load toward LUTs (82\%). Meanwhile, HD2FPGA~\cite{Zhang_HD2FPGA} targets datacenter-class FPGAs (e.g., Xilinx U280), prioritizing automation and maximum throughput over resource minimization.
Ultimately, this analysis reinforces the primary contribution of \Design: while prior works explore varying optimizations such as BRAM elimination, multi-precision support, or maximum throughput, \Design{} excels in providing a highly optimized, application-specific solution with the lowest logic utilization among directly comparable accelerators. This makes \Design{} particularly well-suited for always-on, resource-constrained biomedical applications.

\color{black}

\begin{table*}[!t]
\centering
\vspace{-10pt}
\caption{
%Summary of Related Works on AMS Prediction
\textcolor{black}{Summary of Related Works on AMS Assessment, Prediction, and Hardware-Aware Detection}
}
\label{tab:related_works}
\footnotesize
\setlength{\tabcolsep}{3pt}
\renewcommand{\arraystretch}{1.10}
%\resizebox{\textwidth}{!}{
\begin{tabularx}{\textwidth}{|
>{\centering\arraybackslash}m{1.7cm}|
>{\centering\arraybackslash}m{3.0cm}|
>{\centering\arraybackslash}m{3.0cm}|
>{\centering\arraybackslash}m{1.5cm}|
>{\centering\arraybackslash}m{5.0cm}|
>{\centering\arraybackslash}m{2.55cm}|}
\hline
\textbf{Work} & \textbf{Application} & \textbf{Platform} & \textbf{Dataset} & \textbf{Classifier Accuracy / Diagnostic Metric} & \textbf{Hardware Info. $/$ \newline Efficiency} \\ \hline
Zeng et al. (2024)~\cite{zhengETAL} &
Predict/diagnose AMS from smartwatch SpO$_2$, HR, and 
PI: perfusion index) &
Huawei Smartwatch + Cloud &
$n{=}42$ $>$2500 meters altitude & \textit{Binary Logistic Regression} \newline
SpO$_2$ was an independent predictor of AMS (OR 0.92 per 1\% increase; 95\% CI 0.87–0.97). Accuracy/AUC not reported. & Only session measurement time reported; no hardware efficiency \newline Average measurement per session: 76.39 $\pm$ 12.54 s.
 \\ \hline
\hline
Ye et al. (2023)~\cite{YeETAL} &
Predict AMS from smartwatch-estimated VO$_2$max (plus blood indices) &
Smartwatch + Phone App (Smartwatch Test / Firstbeat); Clinical CPET for reference &
$n{=}46$; low altitude and 3900 meters & \textit{Binary Logistic Regression} \newline
AUC $= 0.785$ using VO$_2$max-SWT alone; AUC $= 0.839$ combining VO$_2$max-SWT + RDW-CV; Specificity up to $88.46\%$ at $29.5$ mL$\cdot$kg$^{-1}\cdot$min$^{-1}$ cutoff; Max.\ F1-measure $= 0.80$ &
software-only  \\ 
\hline
\hline
Yang et al. (2023)~\cite{YangETAL} &
AMS prediction from peripheral blood mononuclear cell (PBMC) for a genetic perspective &
Lab assay (microarray); software model &
Training: 21; Validation: 31 & \textit{SVM-Recursive Feature Elimination (RFE)} \newline
Training day-1 AUC = 1.00 ($N\approx 19$); timeline AUCs: 0.600 (baseline), 0.691 (day-7), 0.673 (day-16) ($N=14$--$21$) &
software-only  \\ 
\hline
\hline
Beidleman et al. (2023)~\cite{BeidlemanETAL} &
AMS incidence \& severity (active vs. passive ascent) &
Field study, 78 soldiers at 3,600 m &
78 participants, 4-day follow-up &  \ \ \textit{
\textcolor{black}{Not prediction model; \ \ \ \ \ \ \ \ \ \ \ \ \ \ \ \ \ \ \ \ \ \ \ \ \ it is included as a field-study reference reporting clinically assessed AMS-C scores, incidence, and severity during active and passive ascent.}
} \newline AMS-C score. Active ascent: 93\% incidence Day 1 vs. 56\% passive; Day 3: 33\% vs. 67\% &
software-only \\ 
\hline
\hline
Wei et al. (2022)~\cite{wei2021using} &
Real-time AMS risk from multi-sensor environment / physiological signals &
Portable SpO$_2$ + Electrocardiogram (ECG) + SensorTag streaming to mobile platform &
n=32 hikers; ~8.4k data points & \ \ \ \ \ \ \ \textit{Bagged Trees} \newline Sensitivity=0.999, Specificity=0.994, Accuracy 0.998, AUC $\approx$ 1.0 (mild-AMS classification)
 &
Used MD-670P Plus oximeter (SpO$_2$, ECG) and TI CC2650 SensorTag (environmental sensors), synchronized via app/PC logging; no hardware efficiency metrics reported \\ 
\hline
\hline
Walzel et al. (2023)~\cite{WalzelETAL} &
Hypoxemia (abnormally low oxygen in the blood--low SpO$_2$) detection &
Apple Watch, Samsung Galaxy Watch, and Withings ScanWatch &
n = 18 (lab hypoxemia protocol, SpO$_2$ 70–100\%) & \ \ \ \ \ \ \ \ \textit{No ML} \newline
Direct comparison with Radical-7 reference oximeter Apple: Acc. 0.93, Sens. 0.91, Spec. 0.95, Samsung: Acc. 0.84, Sens. 0.97, Spec. 0.76, Withings: Acc. 0.89, Sens. 0.92, Spec. 0.86
 &
Commercial smartwatch hardware with built-in proprietary SpO$_2$ sensors
%(wrist-based reflectance photoplethysmography-PPG).
Low-cost, fully integrated; $\leq$4\% error margin vs. Radical-7 reference. \\ 
\hline
\hline
Garg et al. (2021)~\cite{GargETAL} &
Health monitoring \& search-and-rescue framework (wireless sensor network) for mountaineers &
Arduino Mega (ATMega2560) nodes &
System architecture + prototype paper; no dataset usage & \ \ \ \ \ \ \ \ \textit{No ML} \newline
(communication \& power-management framework; emergency operation modes)
 &
Sensors: SpO$_2$/HR (MAX30101), body temp (MAX30205, ±0.1 °C clinical range), ECG (AD8232), GPS (NEO-6M), 9-axis IMU (MPU9250) \\ 
\hline
\hline
This work &
Real-time AMS detection (high-level mobile design) &
Android smartphone + Samsung Galaxy Watch &
Physiological dataset (SpO$_2$, HR, event/time, AMS score) &
 \ \ \textit{HDC Associative Memory (bipolar)} \newline
Acc. $= 0.91$; F1-measure $= 0.90$; decision time $\approx 3.3$ ms (on-phone) &
Smartwatch$\rightarrow$phone pipeline; $\sim$1\% battery drop per session; memory $\approx 92$ B; no cloud dependency \\
\hline
\hline
This work &
Real-time AMS detection (low-level hardware design) &
FPGA (PYNQ-Z2, ARM Cortex-A9) &
Physiological dataset (same as above) &
\ \ \ \ \ \ \ \ \textit{HDC (binary, $D=256$)} \newline
Acc. $= 0.91$ &
LUT $= 1257$, FF $= 1851$, BRAM $= 1$, DSP $= 0$, Power $= 1.6$ W; $16$ clock cycles; $\mathcal{HV}$ generation $= 1$ cycle; $\sim 3.9\times$ lower power vs.\ MLP baseline \\
\hline
\hline
\end{tabularx}
%}
\begin{tablenotes}[flushleft]
\footnotesize
\item[\dag] Hypoxemia vs. AMS: Hypoxemia detection (\( \mathrm{SpO_2} < 90\% \)) is reported as a diagnostic benchmark; it is \emph{not} an AMS classifier.
\item[\ddag] TI SensorTag: Texas Instruments CC2650 multi-sensor BLE (blacktooth Low Energy) module (temperature, pressure, humidity, altitude/barometer).
\item[] \textbf{Abbreviations:} 
AUC — area under the ROC curve; 
VO$_2$max — maximal oxygen consumption; 
RDW-CV — red blood cell distribution width–coefficient of variation; 
CPET — cardiopulmonary exercise test; 
OR — odds ratio; 
CI — confidence interval; 
\(N\), $n$ — sample size; 
PI — perfusion index; 
HRV — heart rate variability
%(ln)RMSSD — natural-log of the root mean square of successive differences; 
%PPG — photoplethysmography; 
%SpO$_2$ — peripheral oxygen saturation.
\vspace{-1.5em}
\end{tablenotes}
\end{table*}

\subsection{AMS Detection Comparison}
\textcolor{black}{Table~\ref{tab:related_works} summarizes prior studies related to AMS assessment, prediction, and monitoring across different platforms, including both model-based AMS prediction works and field studies reporting observed AMS outcomes.}
%perform decision/classifier analysis. 
Most %works 
emphasize the predictive accuracy but overlook hardware efficiency. %of selected biomedical features; 
In contrast, \Design~is designed to address  %our work targets 
both predictive performance and hardware-aware deployment. %hardware efficiency.
In literature, very few existing studies consider on-device computation; instead, many rely on sensor networks or cloud connectivity due to limited processing capability. %few studies are hardware-aware for AMS; many reports are reliant on sensor networks/cloud connectivity due to limited on-device computation.
Zeng et al. explicitly note that ``\textit{The absence of mobile network signals in select regions of the Tibetan Plateau precluded the automated transmission of data to the cloud, necessitating manual data recording in the interim,}”~\cite{zhengETAL}, highlighting the importance of always-on device operation independent of network availability. 

In their study, Zeng et al.~\cite{zhengETAL} employed binary logistic regression with three smartwatch features (SpO\textsubscript{2}, HR, perfusion index) to identify the dominant predictor. Among these, only SpO\textsubscript{2}  showed strong predictive power: %clearly mattered; 
every +1\% increase in SpO\textsubscript{2} corresponded to about an 8\% reduction in AMS risk (e.g., +5\% $\approx$ 34\% lower odds); HR and perfusion index did not provide meaningful additional value. (This corresponds to an odds ratio $\approx$ 0.92 per +1\% SpO\textsubscript{2}; 95\% confidence interval 0.87–0.97; p = 0.001.) They reported only per-session measurement times and lacked hardware efficiency metrics. Similarly, Ye et al.~\cite{YeETAL} applied %a simple statistical classifier, 
binary logistic regression using smartwatch-estimated %, too. Their best model, which utilized 
maximum oxygen consumption 
(VO\textsubscript{2}max) and blood-derived indices such as red blood cell distribution width–coefficient of variation (RDW-CV). Their best model achieved an F1 score of 0.80, but required features (VO\textsubscript{2}max and RDW-CV)  that are more invasive and difficult to obtain compared to SpO\textsubscript{2} and HR. 
%these inputs are more challenging to obtain than SpO\textsubscript{2} and HR. Our AMS-HD achieves an F1 score of 0.84 with wearable/mobile signal processing.
By comparison, \Design~achieves an F1 score of 0.90 using only wearable/mobile physiological signals, while also providing an energy-efficient, hardware-aware implementation suitable for real-time operation on resource-constrained devices.

Yang et al.~\cite{YangETAL}  discuss genetic effects and sample-size considerations for AMS detection. Their SVM-RFE (Support Vector Machine with Recursive Feature Elimination) iteratively removes the least useful features and retrains an SVM to find a small gene set that best separates cases vs. controls. For training day-1, their model perfectly separated severe AMS vs. non-severe AMS (sample size $N\approx 19$). This appears ideal, but with small $N$ and feature selection, likely reflects overfitting/optimism. When the same day-1 model is applied to other time points of the same cohort ($N=14$--$21$ depending on time), performance drops to poor--fair discrimination, consistent with physiological/gene-expression shifts over time and small sample sizes. This underscores the importance of adequate sample size for AMS detection, especially for severe cases, and motivates mobile device designs that enable longitudinal, on-device analysis and future data collection, especially when it comes to sensor-driven data collection and their helps in genetic understanding in the presence of the physiological features.

AMS is also a critical concern for military operations, where on-field testing is often necessary. Beidleman et al.~\cite{BeidlemanETAL} %(2023) 
conducted %an extensive, 
a well-controlled field study involving 78 soldiers at 3,600 m, comparing active ascent (hiking with a backpack) to passive ascent (vehicle transport). \textcolor{black}{They used the measured AMS-C score (a symptom-based index rather than an ML approach) to classify individuals as AMS-susceptible using the Environmental Symptoms Questionnaire (ESQ)~\cite{beidleman2007validation}}. Results showed that active ascenders developed symptoms more rapidly (93\% incidence on day 1 vs.  56\% incidence for passive) but also recovered more quickly (33\% incidence on day 3 vs. 67\% incidence for passive). The overall 4-day incidence and severity did not differ significantly between groups. These findings highlight how ascent profile influences the time course of AMS rather than the overall risk, reinforcing the importance of portable monitoring systems capable of adapting to dynamic field conditions.

Wei et al.~\cite{wei2021using} proposed a comprehensive ML framework for AMS prediction in a sensor-network environment. Their setup combined portable SpO\textsubscript{2} and ECG sensors with a Texas Instruments SensorTag to capture environmental variables such as altitude, temperature, pressure, humidity, and climbing speed. Although the study did not report hardware efficiency, computational complexity, or power analysis, it remains one of the most extensive comparisons of ML classifiers for binary AMS detection. Using over 8,000 data points collected during an 8-hour hike, they evaluated 25 algorithms and found that Bagged Trees achieved the best results with a sensitivity of 0.999, a specificity of 0.994, an accuracy of 0.998, and an AUC close to 1.0. These results demonstrate that multivariate models integrating physiological and environmental signals can achieve very high prediction accuracy. However, the study does not address  %still falls short in emphasizing 
real-time, energy-efficient on-device deployment, which is essential for practical field applications.

Walzel et al.~\cite{WalzelETAL} evaluated the performance of smartwatch-based SpO\textsubscript{2} sensors in a study that provides valuable benchmarking insights. Since our work builds on an %we wanted to maintain the 
open-source Android programming environment, their results are particularly relevant, as they also relied on the Samsung Galaxy Watch for SpO\textsubscript{2} measurements. The study compared three commercial smartwatches (Apple Watch, Samsung Galaxy Watch, and Withings ScanWatch) against a medical-grade reference oximeter (Masimo Radical-7) under a controlled \textcolor{black}{hypoxia} protocol, where participants breathed reduced-oxygen mixtures. Each device 
reported raw SpO\textsubscript{2} values using the manufacturer’s built-in algorithm, which were directly compared with the reference oximeter. Hypoxemia (a decrease in the partial pressure of oxygen--PaO2 in the arterial blood) was defined as SpO\textsubscript{2} $<$90\%, and accuracy was measured as the overall percentage of correct classifications ($<$90\% vs. $\geq$90\%) relative to the reference. Results showed that the Apple Watch outperformed the other devices. While
the reported sensitivity/specificity values reflect %simply 
diagnostic performance metrics rather than outcomes of an ML model, they %but they indicate that Apple outperformed the other devices. This also highlights
underscore the difficulty of achieving high accuracy %that development 
within a purely open-source environment %can be more challenging 
compared to the iOS ecosystem. A study by Rafl et al.~\cite{RaflETAL_Apple} also investigated smartwatch SpO\textsubscript{2} performance with  the Apple Watch Series 6. %shows that 90\% threshold detection is feasible with Apple Watch (notably Series 6); 
In a controlled \textcolor{black}{hypoxia} study involving 24 participants and 642 paired readings, the Apple Watch achieved excellent agreement with a medical-grade oximeter, reporting 0.0\% overall bias and $\approx$1.2\% bias for SpO$_2$ $<$ 90\%. The authors concluded that the device can reliably detect hypoxemia (SpO\textsubscript{2} below 90\%) outside clinical settings. \textcolor{black}{While Apple Watch-based SpO$_2$ monitoring has shown acceptable agreement with medical-grade pulse oximeters under controlled resting or laboratory hypoxemia conditions, its reliability during hiking, military tasks, or other motion-heavy scenarios should be interpreted cautiously~\cite{RaflETAL_Apple,windisch2023accuracy,stove2023assessment}.} Although the Galaxy Watch may fall behind the Apple Watch in performance in some settings, our proposed system still achieves an acceptable decision ratio with a cost-effective approach; nevertheless, %we may benefit from the iOS ecosystem in future work for a more fine-tuned framework design.
future work may benefit from exploring the iOS ecosystem for more fine-tuned framework design.

A notable contribution by  %One-of-a-kind research by 
Garg et al.~\cite{GargETAL}, which is relatively %typically 
scarce in the hardware design literature, presents a system-level platform tailored for mountaineers. The design integrates multiple sensing modules: SpO\textsubscript{2}/HR, body temperature, ECG, GPS, and an inertial measurement unit, all processed on an ATmega2560/Arduino Mega microcontroller. %designed for mountaineers. 
The system also supports data transmission to %authors propose a comprehensive system that also transmits data to 
a master device, enabling team-level monitoring between mountaineers. % (node for a team leader between mountaineers).
\textcolor{black}{While the work demonstrates a mountaineering-oriented, low-temperature-tolerant wireless sensor network design for normal trekking and snow-burial scenarios, the discussion of ML integration remains limited.}
%While the work demonstrates a \textcolor{red}{high-altitude} and snow-resistant design, the discussion of ML integration remains limited. %a stronger ML discussion would be beneficial. 
The authors primarily employ SpO\textsubscript{2} and HR for altitude sickness assessment, which aligns with our objectives. However, instead of relying on an externally powered microcontroller stack, our approach advocates (i) a dedicated hardware pathway for synchronizing these sensors (as in our binary-computing design) or (ii) a high-level mobile platform with smartwatch integration and battery-aware processing that leverages existing on-board sensors (as in the bipolar computing mode of our framework).

Speaking of mobile platforms, Mellor et al.~\cite{Mellor2018SmartphoneHRV} evaluated a smartphone-enabled heart rate variability (HRV) monitor for AMS detection during a 10-day trek to 5,140 $m$ with 21 adults. Their system is not a learning-based design but rather a sensor- and rule-based approach using HRV values. The system records a 55-second HRV sample via a finger sensor linked to a smartphone, which outputs an HRV score ranging from 1 to 100. The study reported weak discrimination for severe AMS using HRV alone, reinforcing our decision to prioritize SpO\textsubscript{2}  as a more reliable predictor. 

Building on these insights, and as summarized 
%Finally, in this work, as 
in the last two rows of Table~\ref{tab:related_works}, our work targets two computing platforms for a more complete framework: (i) a dedicated hardware path for synchronized sensing using binary computing, and (ii) a high-level mobile path that processes smartwatch SpO$_2$/HR signals on-phone with HDC (neuro-symbolic computing) rather than conventional ML. This enables, for the first time, a one-of-its-kind, on-device HDC implementation for AMS detection.

\section{Conclusion}
\label{sec:conclusion}

This work introduced~\Design, the first HDC-based framework for real-time detection of acute mountain sickness (AMS) across embedded and mobile platforms. By leveraging the lightweight and energy-efficient properties of HDC,~\Design~enables fast and reliable AMS classification with minimal computational overhead. A case study on embedded hardware demonstrated low-latency inference and efficient resource utilization, while FPGA-based execution further highlighted the scalability and hardware efficiency of the framework. In addition, implementation on a smartwatch–smartphone platform validated the practicality of deploying~\Design~on mobile devices for real-world use. Importantly, we showed that~\Design~supports both binary and multiclass classification, enabling not only AMS detection but also a finer-grained assessment of its severity. Overall, ~\Design~establishes HDC as a scalable, hardware-aware solution for continuous health monitoring in high-altitude environments, paving the way for broader deployment of real-time, resource-efficient biomedical applications.

\section*{Disclaimer}
The opinions or assertions contained herein are the private views of the authors and are not to be construed as official or reflecting the views of the Army or the Department of War. Any citations of commercial organizations and trade names in this report do not constitute an official Department of the Army endorsement or approval of the products or services of these organizations.

%\vspace{+10pt}

%\section*{Acknowledgment}
%Upon acceptance, the open-source public repository of the generated dataset and the software-hardware code implementations will be shared.

%\newpage
%\balance
\bibliographystyle{IEEEtran}
\bibliography{bibliography, hassan}

@ARTICLE{costanzo_rbme_2022,
  author={Costanzo, Ian and Sen, Devdip and Rhein, Lawrence and Guler, Ulkuhan},
  journal={IEEE Reviews in Biomedical Engineering}, 
  title={Respiratory Monitoring: Current State of the Art and Future Roads}, 
  year={2022},
  volume={15},
  number={},
  pages={103-121},
  doi={10.1109/RBME.2020.3036330}}

@article{rasanen2014modeling,
  title={Modeling dependencies in multiple parallel data streams with hyperdimensional computing},
  author={R{\"a}s{\"a}nen, Okko and Kakouros, Sofoklis},
  journal={IEEE Signal Processing Letters},
  volume={21},
  number={7},
  pages={899--903},
  year={2014},
  publisher={IEEE}
}

@inproceedings{aygun2023linear,
  title={A linear-time, optimization-free, and edge device-compatible hypervector encoding},
  author={Aygun, Sercan and Najafi, M Hassan and Imani, Mohsen},
year={2023},
  booktitle={DATE'23}
}

@article{dunnwald2021use,
  title={The use of pulse oximetry in the assessment of acclimatization to high altitude},
  author={D{\"u}nnwald, Tobias and Kienast, Roland and Niederseer, David and Burtscher, Martin},
  journal={Sensors},
  volume={21},
  number={4},
  pages={1263},
  year={2021},
  publisher={MDPI}
}

@article{wei2021using,
  title={Using machine learning to determine the correlation between physiological and environmental parameters and the induction of acute mountain sickness},
  author={Wei, Chih-Yuan and Chen, Ping-Nan and Lin, Shih-Sung and Huang, Tsai-Wang and Sun, Ling-Chun and Tseng, Chun-Wei and Lin, Ke-Feng},
  journal={BMC bioinformatics},
  volume={22},
  number={Suppl 5},
  pages={628},
  year={2021}
}

@article{wu2020assessment,
  title={Assessment of Acute Mountain Sickness: How to integrate the advantages of the Lake Louise Score and the Chinese AMS Score},
  author={Wu, Yu and Li, Peng and Zhong, Zhifeng and Xie, Jiaxin and Zhou, Simin and Gao, Yuqi and Chen, Jian},
  year={2020}
}

@article{wang2023slow,
  title={A slow feature based LSTM network for susceptibility assessment of acute mountain sickness with heterogeneous data},
  author={Wang, Lei and Xiao, Rong and Chen, Jing and Zhu, Lingling and Shi, Dawei and Wang, Junzheng},
  journal={Biomedical Signal Processing and Control},
  volume={80},
  pages={104355},
  year={2023},
  publisher={Elsevier}
}

@INPROCEEDINGS{wang2024event,
  author={Wang, Lei and Shi, Dawei and Zhu, Lingling and Wang, Junzheng},
  booktitle={2024 39th Youth Academic Annual Conference of Chinese Association of Automation (YAC)}, 
  title={Event-Triggered Pseudo Supervised Meta Learning for Susceptibility Assessment of Acute Mountain Sickness}, 
  year={2024},
  volume={},
  number={},
  pages={1702-1706},
  doi={10.1109/YAC63405.2024.10598519}}

@article{pham2022inflammatory,
  title={Inflammatory gene expression during acute high-altitude exposure},
  author={Pham, Kathy and Frost, Shyleen and Parikh, Keval and Puvvula, Nikhil and Oeung, Britney and Heinrich, Erica C},
  journal={The Journal of physiology},
  volume={600},
  number={18},
  pages={4169--4186},
  year={2022},
  publisher={Wiley Online Library}
}

@article{roach20182018,
  title={The 2018 Lake Louise acute mountain sickness score},
  author={Roach, Robert C and Hackett, Peter H and Oelz, Oswald and B{\"a}rtsch, Peter and Luks, Andrew M and MacInnis, Martin J and Baillie, J Kenneth and Lake Louise AMS Score Consensus Committee},
  journal={High altitude medicine \& biology},
  volume={19},
  number={1},
  pages={4--6},
  year={2018},
  publisher={Mary Ann Liebert, Inc. 140 Huguenot Street, 3rd Floor New Rochelle, NY 10801 USA}
}

@article{hohenhaus1995ventilatory,
  title={Ventilatory and pulmonary vascular response to hypoxia and susceptibility to high altitude pulmonary oedema},
  author={Hohenhaus, E and Paul, A and McCullough, RE and Kucherer, H and Bartsch, P},
  journal={European Respiratory Journal},
  volume={8},
  number={11},
  pages={1825--1833},
  year={1995},
  publisher={Eur Respiratory Soc}
}

@inproceedings{rahimi2016robust,
  title={A robust and energy-efficient classifier using brain-inspired hyperdimensional computing},
  author={Rahimi, Abbas and Kanerva, Pentti and Rabaey, Jan M},
  booktitle={ISLPED},
  pages={64--69},
  year={2016}
}

@inproceedings{shoushtari2024all,
author = {Moghadam, Mehran and Aygun, Sercan and Banitaba, Faeze S. and Najafi, M. Hassan},
title = {{All You Need is Unary: End-to-End Unary Bit-stream Processing in Hyperdimensional Computing}},
year = {2024},
isbn = {9798400706882},
pages = {1–6},
numpages = {6},
location = {Newport Beach, CA, USA},
series = {ISLPED '24}
}

@Article{Kazemi2022,
author={Kazemi, Arman
and M{\"u}ller, Franz
and Sharifi, Mohammad Mehdi
and Errahmouni, Hamza
and Gerlach, Gerald
and K{\"a}mpfe, Thomas
and Imani, Mohsen
and Hu, Xiaobo Sharon
and Niemier, Michael},
title={Achieving software-equivalent accuracy for hyperdimensional computing with ferroelectric-based in-memory computing},
journal={Scientific Reports},
year={2022},
month={Nov},
day={10},
volume={12},
number={1},
pages={19201}
}

@incollection{RAHIMI2020195,
title = {Chapter 8 - Hyperdimensional computing nanosystem: in-memory computing using monolithic 3D integration of RRAM and CNFET},
booktitle = {Memristive Devices for Brain-Inspired Computing},
year = {2020},
author = {Abbas Rahimi and Tony F. Wu and Haitong Li and Jan M. Rabaey and H.-S. Philip Wong and Max M. Shulaker and Subhasish Mitra}
}

@ARTICLE{8327916,
  author={Liu, Siting and Han, Jie},
  journal={IEEE TVLSI}, 
  title={Toward Energy-Efficient Stochastic Circuits Using Parallel Sobol Sequences}, 
  year={2018},
  volume={26},
  number={7},
  pages={1326-1339},
  doi={10.1109/TVLSI.2018.2812214}}

@ARTICLE{Mehran-No-Mult2023,
  author={Moghadam, Mehran Shoushtari and Aygun, Sercan and Najafi, M. Hassan},
  journal={IEEE Embedded Systems Letters}, 
  title={{No-Multiplication Deterministic Hyperdimensional Encoding for Resource-Constrained Devices}}, 
  year={2023},
  volume={15},
  number={4},
  pages={210-213},
  doi={10.1109/LES.2023.3298732}}

@article{kleykoSurvey,
author = {Kleyko, Denis and Rachkovskij, Dmitri and Osipov, Evgeny and Rahimi, Abbas},
title = {A Survey on Hyperdimensional Computing aka Vector Symbolic Architectures, Part II: Applications, Cognitive Models, and Challenges},
year = {2023},
issue_date = {September 2023},
publisher = {Association for Computing Machinery},
address = {New York, NY, USA},
volume = {55},
number = {9},
issn = {0360-0300},
url = {https://doi.org/10.1145/3558000},
doi = {10.1145/3558000}
}

@article{ross2014mutual,
  title={Mutual information between discrete and continuous data sets},
  author={Ross, Brian C},
  journal={PloS one},
  volume={9},
  number={2},
  pages={e87357},
  year={2014},
  publisher={Public Library of Science San Francisco, USA}
}

@article{boos2018relationship,
  title={The relationship between anxiety and acute mountain sickness},
  author={Boos, Christopher J and Bass, Malcolm and O’Hara, John P and Vincent, Emma and Mellor, Adrian and Sevier, Luke and Abdul-Razakq, Humayra and Cooke, Mark and Barlow, Matt and Woods, David R},
  journal={PLoS One},
  volume={13},
  number={6},
  pages={e0197147},
  year={2018},
  publisher={Public Library of Science San Francisco, CA USA}
}

@article{berger2023prevalence,
  title={Prevalence and knowledge about acute mountain sickness in the Western Alps},
  author={Berger, Marc Moritz and H{\"u}sing, Anika and Niessen, Nicolai and Schiefer, Lisa Maria and Schneider, Michael and B{\"a}rtsch, Peter and J{\"o}ckel, Karl-Heinz},
  journal={PLoS One},
  volume={18},
  number={9},
  pages={e0291060},
  year={2023},
  publisher={Public Library of Science San Francisco, CA USA}
}

@article{butson1962generalized,
  title={Generalized hadamard matrices},
  author={Butson, Alton T},
  journal={Proceedings of the American Mathematical Society},
  volume={13},
  number={6},
  pages={894--898},
  year={1962},
  publisher={JSTOR}
}

@article{hedayat1978hadamard,
  title={Hadamard matrices and their applications},
  author={Hedayat, A and Wallis, Walter Dennis},
  journal={The annals of statistics},
  pages={1184--1238},
  year={1978},
  publisher={JSTOR}
}

@article{shlichta1979higher,
  title={Higher-dimensional Hadamard matrices},
  author={Shlichta, P},
  journal={IEEE Transactions on Information Theory},
  volume={25},
  number={5},
  pages={566--572},
  year={1979},
  publisher={IEEE}
}

@incollection{craigen2006hadamard,
  title={Hadamard matrices and Hadamard designs},
  author={Craigen, Robert and Kharaghani, Hadi},
  booktitle={Handbook of combinatorial designs},
  pages={299--305},
  year={2006},
  publisher={Chapman and Hall/CRC}
}

@book{horadam2012hadamard,
  title={Hadamard matrices and their applications},
  author={Horadam, Kathy J},
  year={2012},
  publisher={Princeton university press}
}

@article{pratt1969hadamard,
  title={Hadamard transform image coding},
  author={Pratt, William K and Kane, Julius and Andrews, Harry C},
  journal={Proceedings of the IEEE},
  volume={57},
  number={1},
  pages={58--68},
  year={1969},
  publisher={IEEE}
}

@article{greff2016lstm,
  title={LSTM: A search space odyssey},
  author={Greff, Klaus and Srivastava, Rupesh K and Koutn{\'\i}k, Jan and Steunebrink, Bas R and Schmidhuber, J{\"u}rgen},
  journal={IEEE transactions on neural networks and learning systems},
  volume={28},
  number={10},
  pages={2222--2232},
  year={2016},
  publisher={IEEE}
}

@INPROCEEDINGS{Roodsari_E3HDC_2024,
  author={Roodsari, Mahboobe Sadeghipour and Krautter, Jonas and Meyers, Vincent and Tahoori, Mehdi},
  booktitle={2024 34th International Conference on Field-Programmable Logic and Applications (FPL)}, 
  title={{E3HDC: Energy Efficient Encoding for Hyper-Dimensional Computing on Edge Devices}}, 
  year={2024},
  volume={},
  number={},
  pages={274-280},
  doi={10.1109/FPL64840.2024.00045}}

@ARTICLE{Damiani_MISR_TCAD1990,
  author={Damiani, M. and Olivo, P. and Favalli, M. and Ercolani, S. and Ricco, B.},
  journal={IEEE Transactions on Computer-Aided Design of Integrated Circuits and Systems}, 
  title={{Aliasing in signature analysis testing with multiple input shift registers}}, 
  year={1990},
  volume={9},
  number={12},
  pages={1344-1353},
  doi={10.1109/43.62779}}

@ARTICLE{aygun2024sobol,
  author={Aygun, Sercan and Hassan Najafi, M.},
  journal={IEEE Transactions on Computer-Aided Design of Integrated Circuits and Systems}, 
  title={Sobol Sequence Optimization for Hardware-Efficient Vector Symbolic Architectures}, 
  year={2025},
  volume={44},
  number={3},
  pages={937-950},
  doi={10.1109/TCAD.2024.3463544}}

@inproceedings{masum2025fly,
  author    = {Abu Kaisar Mohammad Masum and Mehran Shoushtari Moghadam and Sabrina Hassan Moon and Ahmed Mamdouh Mohamed Ahmed and M. Hassan Najafi and Dayane Reis and Sercan Aygun},
  title     = {{On-the-Fly Hadamard Hypervector Processing for Efficient Hyperdimensional Computing}},
  booktitle = {Design Automation Conference (DAC)},
  year      = {2025}
}

@inproceedings{aygun2024uhd,
  title={{uHD: Unary processing for lightweight and dynamic hyperdimensional computing}},
  author={Aygun, Sercan and Moghadam, Mehran Shoushtari and Najafi, M Hassan},
  booktitle={2024 Design, Automation \& Test in Europe Conference \& Exhibition (DATE)},
  pages={1--6},
  year={2024},
  organization={IEEE}
}

@ARTICLE{Najafi_TVLSI2018_Unary,
  author={Najafi, M. Hassan and Lilja, David. J. and Riedel, Marc D. and Bazargan, Kia},
  journal={IEEE Transactions on Very Large Scale Integration (VLSI) Systems}, 
  title={Low-Cost Sorting Network Circuits Using Unary Processing}, 
  year={2018},
  volume={26},
  number={8},
  pages={1471-1480},
  doi={10.1109/TVLSI.2018.2822300}}

@article{RahimiHamming,
author = {Schmuck, Manuel and Benini, Luca and Rahimi, Abbas},
title = {Hardware Optimizations of Dense Binary Hyperdimensional Computing: Rematerialization of Hypervectors, Binarized Bundling, and Combinational Associative Memory},
year = {2019},
volume = {15},
number = {4},
journal = {J. Emerg. Technol. Comput. Syst.},
month = oct,
articleno = {32},
numpages = {25}
}

@INPROCEEDINGS{Roodsari_OTFGEnc_DATE24,
  author={Roodsari, Mahboobe Sadeghipour and Krautter, Jonas and Tahoori, Mehdi},
  booktitle={2024 Design, Automation \& Test in Europe Conference \& Exhibition (DATE)}, 
  title={{OTFGEncoder - HDC: Hardware-efficient Encoding Techniques for Hyperdimensional Computing}}, 
  year={2024},
  volume={},
  number={},
  pages={1-2},
  doi={10.23919/DATE58400.2024.10546523}}

@Article{YeETAL,
author="Ye, Xiaowei
and Sun, Mengjia
and Yu, Shiyong
and Yang, Jie
and Liu, Zhen
and Lv, Hailin
and Wu, Boji
and He, Jingyu
and Wang, Xuhong
and Huang, Lan",
title="Smartwatch-Based Maximum Oxygen Consumption Measurement for Predicting Acute Mountain Sickness: Diagnostic Accuracy Evaluation Study",
journal="JMIR Mhealth Uhealth",
year="2023",
month="Jul",
day="6",
volume="11",
pages="e43340",
issn="2291-5222",
doi="10.2196/43340",
url="https://mhealth.jmir.org/2023/1/e43340",
url="https://doi.org/10.2196/43340",
url="http://www.ncbi.nlm.nih.gov/pubmed/37410528"
}

@article{zhengETAL,
  author       = {Zeng, Zhengyang and Li, Lili and Hu, Li'ao and Wang, Kang and Li, Lun},
  title        = {Smartwatch measurement of blood oxygen saturation for predicting acute mountain sickness: Diagnostic accuracy and reliability},
  journaltitle = {Digital Health},
  date         = {2024-09-27},
  year         = {2024},
  volume       = {10},
  eid          = {20552076241284910},
  pages        = {20552076241284910},
  doi          = {10.1177/20552076241284910},
  issn         = {2055-2076},
  eprinttype   = {pubmed},
  eprint       = {39351311},
  pmid         = {39351311},
  note         = {PMCID: PMC11440541; eCollection 2024 Jan--Dec},
  langid       = {english}
}

@article{YangETAL,
  author       = {Yang, Min and Wu, Yang and Yang, Xing-biao and Liu, Tao and Zhang, Ya and Zhuo, Yue and Luo, Yong and Zhang, Nan},
  title        = {Establishing a prediction model of severe acute mountain sickness using machine learning of support vector machine recursive feature elimination},
  journaltitle = {Scientific Reports},
  date         = {2023-03-21},
  year         = {2023},
  volume       = {13},
  number       = {1},
  eid          = {4633},
  pages        = {4633},
  doi          = {10.1038/s41598-023-31797-0},
  url          = {https://doi.org/10.1038/s41598-023-31797-0},
  issn         = {2045-2322},
  langid       = {english}
}

@article{BeidlemanETAL,
  author       = {Beidleman, Beth A. and Figueiredo, Peter S. and Landspurg, Steven D. and Femling, Jon K. and Williams, Jason D. and Staab, Janet E. and Buller, Mark J. and Karl, J. Philip and Reilly, Aaron J. and Mayschak, Trevor J. and Atkinson, Emma Y. and Mesite, Timothy J. and Hoyt, Reed W.},
  title        = {Active ascent accelerates the time course but not the overall incidence and severity of acute mountain sickness at 3,600 m},
  journaltitle = {Journal of Applied Physiology},
  date         = {2023-08-01},
  year         = {2023},
  volume       = {135},
  number       = {2},
  pages        = {436--444},
  doi          = {10.1152/japplphysiol.00216.2023},
  url          = {https://doi.org/10.1152/japplphysiol.00216.2023},
  issn         = {8750-7587},
  eissn        = {1522-1601},
  eprinttype   = {pubmed},
  eprint       = {37318986},
  pmid         = {37318986},
  note         = {Epub 2023-06-15; PMCID: PMC10538982},
  langid       = {english}
}

@article{li2025acute,
  title={Acute mountain sickness prediction: a concerto of multidimensional phenotypic data and machine learning strategies in the framework of predictive, preventive, and personalized medicine},
  author={Li, Wenhui and Zhang, Meng and Hu, Yangyi and Shen, Pan and Bai, Zhijie and Huangfu, Chaoji and Ni, Zhexin and Sun, Dezhi and Wang, Ningning and Zhang, Pengfei and others},
  journal={EPMA Journal},
  pages={1--20},
  year={2025},
  publisher={Springer}
}

@article{wang2024recent,
  title={Recent advances in predicting acute mountain sickness: from multidimensional cohort studies to cutting-edge model applications},
  author={Wang, Boyuan and Chen, Shanji and Song, Jinfeng and Huang, Dan and Xiao, Gexin},
  journal={Frontiers in Physiology},
  volume={15},
  pages={1397280},
  year={2024},
  publisher={Frontiers Media SA}
}

@article{WalzelETAL,
  author       = {Walzel, Simon and Mikus, Radek and Rafl-Huttova, Veronika and Rozanek, Martin and Bachman, Thomas E. and Rafl, Jakub},
  title        = {Evaluation of Leading Smartwatches for the Detection of Hypoxemia: Comparison to Reference Oximeter},
  journaltitle = {Sensors},
  date         = {2023-11-14},
  year         = {2023},
  volume       = {23},
  number       = {22},
  pages        = {9164},
  eid          = {9164},
  issn         = {1424-8220},
  eprinttype   = {pubmed},
  eprint       = {38005550},
  pmid         = {38005550},
  note         = {PMCID: PMC10674783},
  langid       = {english}
}

@article{RaflETAL_Apple,
author = {Jakub Rafl and Thomas E Bachman and Veronika Rafl-Huttova and Simon Walzel and Martin Rozanek},
title ={Commercial smartwatch with pulse oximeter detects short-time hypoxemia as well as standard medical-grade device: Validation study},
journal = {DIGITAL HEALTH},
volume = {8},
number = {},
pages = {20552076221132127},
year = {2022}
}

@article{GargETAL,
title = {Healthcare monitoring of mountaineers by low power Wireless Sensor Networks},
journal = {Informatics in Medicine Unlocked},
volume = {27},
pages = {100775},
year = {2021},
issn = {2352-9148},
author = {Rajesh Kumar Garg and Jyoti Bhola and Surender Kumar Soni}
}

@article{Mellor2018SmartphoneHRV,
  author    = {Mellor, Adrian M. B. and Bakker-Dyos, Josh and O'Hara, John and Woods, David Richard and Holdsworth, David A. and Boos, Christopher J.},
  title     = {Smartphone-Enabled Heart Rate Variability and Acute Mountain Sickness},
  journal   = {Clinical Journal of Sport Medicine},
  volume    = {28},
  number    = {1},
  pages     = {76--81},
  year      = {2018},
  month     = jan,
  doi       = {10.1097/JSM.0000000000000427}
}

@INPROCEEDINGS{Zhang_HD2FPGA,
  author={Zhang, Tinaqi and Salamat, Sahand and Khaleghi, Behnam and Morris, Justin and Aksanli, Baris and Rosing, Tajana Simunic},
  booktitle={2023 24th International Symposium on Quality Electronic Design (ISQED)}, 
  title={HD2FPGA: Automated Framework for Accelerating Hyperdimensional Computing on FPGAs}, 
  year={2023}
  }

@inproceedings{Salamat_F5HD,
author = {Salamat, Sahand and Imani, Mohsen and Khaleghi, Behnam and Rosing, Tajana},
title = {F5-HD: Fast Flexible FPGA-based Framework for Refreshing Hyperdimensional Computing},
year = {2019},
booktitle = {Proceedings of the 2019 ACM/SIGDA International Symposium on Field-Programmable Gate Arrays},
pages = {53–62}
}

@ARTICLE{Imani_QuantHD,
  author={Imani, Mohsen and Bosch, Samuel and Datta, Sohum and Ramakrishna, Sharadhi and Salamat, Sahand and Rabaey, Jan M. and Rosing, Tajana},
  journal={IEEE Transactions on Computer-Aided Design of Integrated Circuits and Systems}, 
  title={QuantHD: A Quantization Framework for Hyperdimensional Computing}, 
  year={2020},
  volume={39},
  number={10},
  pages={2268-2278},
  doi={10.1109/TCAD.2019.2954472}}

@inproceedings{Moghadam_Robust_GLSVLSI25,
author = {Moghadam, Mehran and Masum, Abu Kaisar Mohammad and Aygun, Sercan and Najafi, M. Hassan},
title = {{Robust Data Processing for Vector Symbolic Computing}},
year = {2025},
isbn = {9798400714962},
publisher = {Association for Computing Machinery},
address = {New York, NY, USA},
url = {https://doi.org/10.1145/3716368.3735287},
doi = {10.1145/3716368.3735287},
booktitle = {Proceedings of the Great Lakes Symposium on VLSI 2025},
pages = {823–828},
numpages = {6},
location = {
},
series = {GLSVLSI '25}
}

@article{beidleman2017predicting,
  title={Predicting individual risk of altitude illness using real-time monitoring of accumulated hypoxic debt},
  author={Beidleman, Beth and Welles, Alexander and Buller, Mark},
  journal={Journal of Science and Medicine in Sport},
  volume={20},
  pages={S94--S95},
  year={2017},
  publisher={Elsevier}
}

@article{stove2023assessment,
  author  = {St{\o}ve, Martin P. and Graversen, Anne H. and S{\o}rensen, Jens},
  title   = {Assessment of Noninvasive Oxygen Saturation in Patients With {COPD} During Pulmonary Rehabilitation: Smartwatch versus Pulse Oximeter},
  journal = {Respiratory Care},
  year    = {2023},
  volume  = {68},
  number  = {8},
  pages   = {1041--1048},
  doi     = {10.4187/respcare.10760},
  pmid    = {37193599},
  pmcid   = {PMC10353168}
}

@article{windisch2023accuracy,
  author  = {Windisch, Paul and Schr{\"o}der, Christoph and F{\"o}rster, Rebecca and Cihoric, Nata{\v{s}}a and Zwahlen, Daniel R.},
  title   = {Accuracy of the {Apple Watch} Oxygen Saturation Measurement in Adults: A Systematic Review},
  journal = {Cureus},
  year    = {2023},
  volume  = {15},
  number  = {2},
  pages   = {e35355},
  doi     = {10.7759/cureus.35355},
  pmid    = {36974257},
  pmcid   = {PMC10039641}
}

@article{beidleman2007validation,
  author  = {Beidleman, Beth A. and Muza, Stephen R. and Fulco, Charles S. and Rock, Paul B. and Cymerman, Allen},
  title   = {Validation of a Shortened Electronic Version of the Environmental Symptoms Questionnaire},
  journal = {High Altitude Medicine \& Biology},
  year    = {2007},
  volume  = {8},
  number  = {3},
  pages   = {192--199},
  doi     = {10.1089/ham.2007.1016},
  pmid    = {17824819}
}

@ARTICLE{Seizure_HDC_TBME_2020,
  author={Burrello, Alessio and Schindler, Kaspar and Benini, Luca and Rahimi, Abbas},
  journal={IEEE Transactions on Biomedical Engineering}, 
  title={Hyperdimensional Computing With Local Binary Patterns: One-Shot Learning of Seizure Onset and Identification of Ictogenic Brain Regions Using Short-Time iEEG Recordings}, 
  year={2020},
  volume={67},
  number={2},
  pages={601-613},
  doi={10.1109/TBME.2019.2919137}}

@ARTICLE{Sleep_Apnea_HDC_TBME_2024,
  author={Chen, Tian and Zhang, Jingtao and Xu, Zeju and Redmond, Stephen J. and Lovell, Nigel H. and Liu, Guanzheng and Wang, Changhong},
  journal={IEEE Transactions on Biomedical Engineering}, 
  title={Energy-Efficient Sleep Apnea Detection Using a Hyperdimensional Computing Framework Based on Wearable Bracelet Photoplethysmography}, 
  year={2024},
  volume={71},
  number={8},
  pages={2483-2494},
  doi={10.1109/TBME.2024.3377270}}

@ARTICLE{Phoneme_Recog_HDC_TBME_2024,
  author={Ni, Yang and Yang, Ye and Chen, Hanning and Wang, Xianhui and Lesica, Nicholas and Zeng, Fan-gang and Imani, Mohsen},
  journal={IEEE Transactions on Biomedical Engineering}, 
  title={Hyperdimensional Brain-Inspired Learning for Phoneme Recognition With Large-Scale Inferior Colliculus Neural Activities}, 
  year={2024},
  volume={71},
  number={11},
  pages={3098-3110},
  doi={10.1109/TBME.2024.3408279}}

@ARTICLE{ID_VSA_Moghadam_TVLSI2026,
  author={Moghadam, Mehran and Masum, Abu and Aygun, Sercan and Najafi, M. Hassan},
  journal={IEEE Transactions on Very Large Scale Integration (VLSI) Systems}, 
  title={{Independent and Dynamic Vector Symbolic Architecture for Hardware-Efficient Edge AI}}, 
  year={2026},
  volume={34},
  number={4},
  pages={1110-1123},
  doi={10.1109/TVLSI.2026.3671174}}

@String{Computing = "Computing" }

@String{Academic = "Academic Press" }

@String{AMS = "American Mathematical Society" }

@String{Springer = "Springer-Verlag" }

@ArtifactSoftware{R,
    title = {R: A Language and Environment for Statistical Computing},
    author = {{R Core Team}},
    organization = {R Foundation for Statistical Computing},
    address = {Vienna, Austria},
    year = {2019},
    url = {https://www.R-project.org/},
}

@ARTICLE{Najafi_TVLSI_2019,
  author={Najafi, M. Hassan and Jenson, Devon and Lilja, David J. and Riedel, Marc D.},
  journal={IEEE Transactions on Very Large Scale Integration (VLSI) Systems}, 
  title={Performing Stochastic Computation Deterministically}, 
  year={2019},
  volume={27},
  number={12},
  doi={10.1109/TVLSI.2019.2929354}}

@article{kanerva2009hyperdimensional,
  title={Hyperdimensional computing: An introduction to computing in distributed representation with high-dimensional random vectors},
  author={Kanerva, Pentti},
  journal={Cognitive Computation},
  volume={1},
  number={2},
  pages={139--159},
  year={2009},
  publisher={Springer}
}

@inproceedings{imani2019framework,
  title={A framework for collaborative learning in secure high-dimensional space},
  author={Mohsen Imani and Kim, Yeseong and Riazi, Sadegh and Messerly, John and Liu, Patric and Koushanfar, Farinaz and Rosing, Tajana},
  booktitle={IEEE CLOUD},
  pages={435--446},
  year={2019}
}

@inproceedings{10.1145/3649476.3658795,
author = {Ayar, Alaaddin Goktug and Aygun, Sercan and Najafi, M. Hassan and Margala, Martin},
title = {Word2HyperVec: From Word Embeddings to Hypervectors for Hyperdimensional Computing},
year = {2024},
booktitle = {Great Lakes Symposium on VLSI},
pages = {355–356},
numpages = {2},
location = {Clearwater, FL, USA}
}

\end{document}